\documentstyle[version2,preprint,aps,epsf]{revtex} 
\begin{document}
\begin{title}
Electronic Raman Scattering in Superconductors as a Probe of\\
Anisotropic Electron Pairing
\end{title}
\author{T. P. Devereaux$^{1}$ and D. Einzel$^{2}$}
\begin{instit}
$^{1}$Department of Physics\\
University of California\\
Davis, CA 95616\\
\end{instit}
\begin{instit}
$^{2}$Walther--Meissner--Institut f\"ur Tieftemperaturforschung\\
D--85748 Garching\\
Federal Republic of Germany\\
\end{instit}

\begin{abstract}
A gauge invariant
theory for electronic Raman scattering for superconductors
with anisotropic pairing symmetry is analyzed in detail. It is shown that Raman
scattering in anisotropic superconductors
provides a wealth of polarization-dependent information
which probes the detailed
angular dependence of the superconducting ground state order
parameter. The Raman spectra shows a unique polarization dependence
for various anisotropic pair--state symmetries which affects the 
peak position of the
spectra and generates symmetry dependent
low frequency and temperature power--laws which can be used
to uniquely identify the magnitude and symmetry of the energy gap.
In particular, we calculate the collective modes and the subsequent
symmetry--dependent Raman spectra for a $d_{x^{2}-y^{2}}$ superconductor and
compare our results to the relevant data on the
cuprate systems as well as theoretical predictions for $s$--wave, anisotropic
$s$--wave and $s+id$ energy gaps. Favorable agreement is shown with the 
predictions for
$d_{x^{2}-y^{2}}$ pairing and the experimental data on
YBa$_{2}$Cu$_{3}$O$_{7}$, Bi$_{2}$Sr$_{2}$CaCu$_{2}$O$_{8}$ and
Tl$_{2}$Ba$_{2}$CuO$_{6}$. 
\end{abstract}
\receipt{\today}
\narrowtext

\section{Introduction}

        Knowledge of the symmetry of the energy gap in
superconductors provides a major step towards unraveling the puzzle of
superconductivity in unconventional superconductors. While the evidence
continues to accrue, the
identification of the pair state in the heavy--fermion and cuprate systems
has proven to be somewhat elusive \cite{major}.  While recent photoemission
experiments \cite{shen} do allow for an
angular--dependent determination of the gap 
and Josephson tunneling measurements
have probed the phase of the gap \cite{phase}, by far the most abundant
information that has led to speculation of a non--BCS ground state has been
focussed on the presence of power--laws in the low frequency and/or
temperature behavior
of transport and thermodynamic quantities \cite{expmt}.
However, due to the averaging over
the entire Fermi surface, it is well known that the
power--laws themselves do not uniquely identify the ground state
symmetry of the order parameter but only can give the topology of the nodes
of the energy gap along the Fermi surface, e.g., whether the gap vanishes
on points and/or lines on the Fermi surface. Thus one cannot distinguish 
between
different representations of the energy gap which have the same topology.
For instance, for the case of $d$--wave tetragonal superconductors, there 
are five pure representations which have line nodes on the Fermi surface.
Further, the energy gap can become smeared due to 
inelastic quasiparticle collisions,
making a fully gapped superconductor seem like a one
with gap nodes. While in principle the latter effect can be minimized by
limiting the experiment to very low temperatures, the two-particle correlation
functions determining the density, spin or current responses
do not have the freedom to probe various portions of the gap around
the Fermi surface, presenting a fundamental obstacle to uniquely
identifying the pair state representation for the superconductor.

        However, it is well known that Raman
scattering has the ability to measure various degrees of freedom by
simply rotating the incident and scattered photon polarization
orientations.
Formally, Raman fluctuations may be viewed as anisotropic mass fluctuations
around the Fermi surface which do not obey a conservation law \cite{kost},
such as, e.g., density fluctuations.  While the isotropic
density fluctuations between unit cells will be screened due to 
their coupling to
long--range Coulomb forces, the intracell mass fluctuations can be
anisotropic around the Fermi surface with no net charge and thus
can be unscreened, providing a scattering mechanism for incoming
photons.

The intensity of scattered light in a Raman experiment can be written in terms 
of a differential photon scattering cross section 
$\partial^2\sigma/\partial\omega\partial\Omega$ as
\begin{eqnarray}
\frac{\partial^2\sigma}{\partial\omega\partial\Omega}&=&
\frac{\omega_{\rm S}}{\omega_{\rm I}}\ r_0^2\ S_{\gamma\gamma}({\bf q},\omega),
\nonumber \\
S_{\gamma\gamma}({\bf q},\omega)&=&-\frac{1}{\pi}
\left[1+n(\omega)\right]{\rm Im}\ \chi_{\gamma\gamma}({\bf q},\omega).
\end{eqnarray}
Here $r_0=e^2/mc^2$ is the Thompson radius, $\omega_{\rm I}$, $\omega_{\rm S}$
is the frequency of the incoming and scattered photon, respectively, and
we have set $\hbar=k_{B}=1$.
$S_{\gamma\gamma}$ is the generalized structure function, which is related to 
the imaginary part to the Raman response function $\chi_{\gamma\gamma}$
through the fluctuation--dissipation theorem, the second part of Eq. (1).
Finally $n(\omega)=1/[\exp( \omega/T)-1]$ is the Bose--Einstein 
distribution function. 
The generalization of the usual density--density correlation ($S$) and response
($\chi$) functions to the case of Raman scattering can be generated by 
weighting the sums over momenta ${\bf k}$ with the square of the Raman 
vertex $\gamma({\bf k})$.

For small momentum transfers and incident light energies smaller than
the optical band gap (non--resonant scattering), 
the vertex for Raman scattering can be written in 
terms of the curvature of the energy band dispersion
$\epsilon({\bf k})$,
\begin{equation}
\gamma({\bf k}) = m \sum_{\alpha,\beta}e_{\alpha}^{\rm S}
{\partial^2\epsilon({\bf  k})\over{\partial k_{\alpha}\partial
k_{\beta}}} e_{\beta}^{\rm I},
\end{equation}
where ${\bf  \hat e^{S,I}}$ denote the scattered and incident
polarization light vectors, respectively, which select elements of the
Raman tensor.  The symmetry of the underlying
crystal can be taken into account by expanding $\gamma$
in terms of a complete set of crystal harmonics $\phi_{L}$ defined on the Fermi
surface, i.e., \cite{allen}
\begin{equation}
\gamma({\bf k})=\sum_{L,m}\gamma_{L}^{m}\phi_{L}^{m}({\bf k}),
\end{equation}
where the index $L$ represents the $L$--th order contribution to the vertex
which transforms according to the $m$--th irreducible representation of the
point group symmetry of the crystal.  The quantum numbers $L, m$
classify the anisotropy of the 
Raman fluctuations around the
Fermi surface. The full $k$--dependence is thus described
by the addition of the basis functions
(which become progressively more anisotropic for higher $L$ )
with different weights $\gamma_{L}^{m}$.
While the charge, spin, and current density response probes only a single
$L$ channel ($\gamma({\bf k})=1, L=0$, for the charge and spin density, while
$\gamma({\bf k})={\bf k}, L=1$, for the current density),
in principle all even $L$ channels
can contribute to the Raman vertex for bands which are non--parabolic. Thus
by choosing the polarization light vectors accordingly, one can select
different $L, m$ channels which allow for different projections onto
the Fermi surface. Therefore, electronic Raman scattering can provide detailed
information which allows insight to the angular dependence
of the pair--state symmetry.

        In accord with this fact,
the experimental results on the cuprate systems reveal a wealth
of polarization--dependent information that provides
detailed evidence for determining
the actual symmetry of the gap \cite{us}. The existing
body of data on the cuprate systems \cite{review}
reveal five main points: 1) in contrast to
conventional superconductors such as Nb$_{3}$Sn, no clear well--defined
gap is seen for any polarization orientation even at the lowest
temperatures measured ($T/T_{c}=0.03$) \cite{cooper1},
2) the peak of the
spectrum lies roughly at 30\% higher frequency shifts
for the polarization orientation which selects
B$_{1g}$ symmetry compared to all other symmetries [10-13],
3) there are indications that the
temperature dependence of the peak in the B$_{1g}$ channel follows more
closely to a BCS form than any other symmetry \cite{hackl1},
4) the low frequency
Raman shifts vary roughly as $\omega^{3}$ for B$_{1g}$ symmetries and
linearly in $\omega$ for the others [9-13], and 5) the ratio of
residual scattering in the superconducting state to the normal state
is smallest for the B$_{1g}$ case compared to all other
configurations \cite{hackl3}.
Such a rich spectrum of information should provide a
stringent test for various candidates of the pairing symmetry states 
\cite{phonons}.

        The purpose of the present paper is to investigate the
polarization dependence of the Raman spectra for a superconductor
in the weak coupling limit with anisotropic pairing symmetry.
We calculate
the electronic Raman scattering for a tetragonal superconductor at
finite temperatures and for various polarization orientations in a
gauge invariant manner and find a rich polarization
dependence of the spectra that can be used to uniquely identify the energy
gap. In particular, we examine the Raman response for $s$--, $d$--, $s+id$--, 
and anisotropic $s$--wave superconductors and find that a $d_{x^{2}-y^{2}}$
state agrees surprising well with the current
information on electronic Raman scattering in 
YBa$_{2}$Cu$_{3}$O$_{7}$, Bi$_{2}$Sr$_{2}$CaCu$_{2}$O$_{8}$ and
Tl$_{2}$Ba$_{2}$CuO$_{6}$. 

        The plan of the paper is as follows: The ground work for the gauge
invariant theory of
electronic Raman scattering in anisotropic superconductors is reviewed in
Section 2 using a kinetic equation approach.
Section 3 concerns the connection of band structure to the Raman vertex
and its relation to Fermi surfaces. Section 4 gives
our results for the Raman response evaluated on a cylindrical Fermi surface 
for four types of energy gaps,
i) $s$--wave, ii) $d$--wave, iii) $s+id$--wave, and iv)
a special type of anisotropic $s$--wave.
Section 5 presents a comparison of the theory to the data on three cuprate
systems and contains our conclusions. 
Appendix A deals with a solution of the weak coupling gap equation for the 
case of anisotropic gaps, Appendix B contains details of the
theory for the case of (elastic or inelastic) quasiparticle scattering 
in the normal state, while lastly Appendix C
is devoted to our calculations for the massive
collective modes and the subsequent role of vertex corrections
using a diagrammatic approach. A brief account of our
work has recently appeared \cite{us}.

\section{Kinetic theory of electronic Raman scattering in
anisotropic superconductors}
\subsection{Formalism}
In this section we describe a kinetic equation approach for
calculating generalized gauge--invariant response functions in anisotropic
superconductors, with the majority of our attention being focused
on the electronic Raman response. Other formulations of the calculation
are possible, and in particular a diagrammatic approach for the 
gauge invariant Raman response is discussed in detail in Appendix C.
Since this approach has been discussed before by us \cite{us}, we will
concentrate on the kinetic equation approach here.

We consider an anisotropic superconductor in which the electronic
states are characterized by a momentum $ {\bf k}$, an energy (band) dispersion
$\epsilon_k=\mu+\xi_k$ (with $\mu$ the Fermi energy), a (band) group velocity 
${\bf v}_k=\nabla_k\epsilon_k$, an inverse effective mass tensor
$M_{i j}^{-1}
({\bf k})=\partial^2\epsilon_k/\partial k_i\partial k_j$,
an energy gap $\Delta_k$ and excitation
energies $E_k=[\xi_k^2+|\Delta_k|^2]^{1/2}$. In global thermodynamic
equilibrium such a system is described by a diagonal equilibrium phase space
distribution function $n_k^0$
\begin{equation}
n_k^0=<c_k^{\dagger}c_k>=
u_k^2f(E_k)+v_k^2[1-f(E_k)]=\frac{1}{2}\left[1-\xi_k\theta_k\right] \ \ ,
\end{equation}
with $\theta_k=(1/2E_k)\tanh(E_k/2T)$, $f(E_k)$ the Fermi function taken
at the Bogoliubov quasiparticle energy $E_k$ and the usual coherence factors
$u_k^2=\frac{1}{2}[1+\xi_k/E_k]$ and $v_k^2=\frac{1}{2}[1-\xi_k/E_k]$
for particle--like and hole--like Bogoliubov quasiparticles. 
In a superconductor there exists in addition to the diagonal average
$n_k^0$ the off--diagonal average $g_k^0$
\begin{equation}
g_k^0=<c_{-k}c_k>=-\Delta_k\theta_k .
\end{equation}
The superconducting equilibrium energy gap is then determined from the
self consistency equation 
\begin{equation} 
\Delta_k=\sum_{\bf p} V_{k p} \ g_p^0 ,
\end{equation} 
with $V_{k p}=-|V_{k p}|$ the pairing interaction. 

Such a system is assumed to be subject to external perturbation potentials
$U_k^{\rm\ ext}({\bf q},\omega)\propto \exp(i{\bf q}\cdot{\bf r}-i\omega t)$ 
which are generated in the usual way by an expansion of the Hamiltonian 
$\propto ({\bf p}-e{\bf A}^{\rm\ ext}/c)^2/2m+e\Phi^{\rm\ ext}$
containing ${\bf q}$ and $\omega$ dependent 
scalar and vector electromagnetic potentials $\Phi^{\rm\ ext}
({\bf q},\omega)$ and ${\bf A}^{\rm\ ext}({\bf q},\omega)$, respectively,  
to second order in the vector potential 
(including ${\bf p}\cdot{\bf A}$--terms):
\begin{eqnarray}
U_k^{\rm\ ext}&=&e\Phi^{\rm\ ext} 
+{\bf v}_k\cdot\left(-\frac{e}{c}{\bf A}^{\rm\ ext}\right)
+\frac{e^2}{c^2}\ A_i^{\rm I}\cdot M^{-1}_{i j}({\bf k})
\cdot A_j^{\rm S}, \nonumber \\
&=&e\Phi^{\rm ext}+{\bf v}_k\cdot\delta\vec{\epsilon}_1+
\gamma({\bf k})\delta\epsilon_{\gamma},
\end{eqnarray}
where
\begin{eqnarray}
\delta\vec{\epsilon}_1&=&-\frac{e}{c}{\bf A}^{\rm ext}, \nonumber \\
\delta\epsilon_{\gamma}&=&r_0\ |{\bf A}^{\rm I}|\ |{\bf A}^{\rm S}|, \nonumber
\end{eqnarray}
$\gamma({\bf k})$ denotes the Raman vertex given by Eq. (2).
The last term in Eq. (7) can be
interpreted to describe the scattering of an incident photon 
of frequency $\omega_{\rm I}$ represented by the vector potential 
${\bf A}^{\rm I}$, into electronic excitations such as 
particle--hole, Bogoliubov quasiparticle or magnon pairs and an
outgoing photon (Stokes process) of frequency 
$\omega_{\rm S}=\omega_{\rm I}-\omega$ 
(vector potential ${\bf A}^{\rm S}$) with an associated
total momentum transfer $ {\bf q}$. 

The (linear) response  of the distribution function $\delta
n_k({\bf q},\omega)=n_k({\bf q},\omega)-n_k^0$ 
in the absence of dissipation is given by the solution of
the collisionless (particle--hole--symmetric) kinetic equation 
\cite{B-MN,wolfle},   
\begin{equation}  
\delta n_k=\frac{\eta_k}{\omega-\eta_k}(\phi_k-\lambda_k)\delta\epsilon_k
-\lambda_k\delta\epsilon_k^+
+\lambda_k(\omega+\eta_k)\frac{\delta\Delta_k^-}{2|\Delta_k|} \ \ ,
\end{equation}
where $\eta_k \equiv {\bf v}_k \cdot {\bf q}$, 
\begin{equation}
\phi_k= -\frac{\partial n_k^0}{\partial \xi_k} = 
\frac{|\Delta_k|^2}{E_k^2} \theta_k +\frac{\xi_k^2}{E_k^2} \varphi_k \ \ , 
\end{equation}
and $\varphi_k=-\partial f(E_k)/\partial E_k$. In addition we have defined
\begin{equation}
\delta\Delta_k^s=\frac{\delta\Delta_k\Delta_k^{\dagger}
+s\Delta_k\delta\Delta_k^{\dagger}}{2{|\Delta_k|}} \ \ \ ; \ \ \ s=\pm 1.
\end{equation}
The cases $s=-1$ and $s=+1$ distinguish the coupling of the response of the 
pair--correlated electron system to phase and amplitude fluctuations 
of the order parameter, respectively, and will be discussed later. 
The quantity $\lambda_k$
is the pair response function introduced by Tsuneto
\cite{Tsuneto1} which, in the limit 
$q\xi \ll 1$, with $\xi$ the coherence length,
is given by,  
\begin{equation} 
\lambda_k({\bf q},\omega)=-4|\Delta_k|^2
\frac{(\omega^2-\eta_k^2)\theta_k+\eta_k^2\Phi_k}
{\omega^2(\omega^2-4E_k^2)-\eta_k^2(\omega^2-4\xi_k^2)} \ \ .
\end{equation}
A more general expression for of $\lambda_k$ valid also for 
$\xi^{-1}<q\ll k_{\rm F}$ is given in Ref. \cite{Hirschfeld1}. 
The total {\it diagonal} quasiparticle energy shift $\delta\epsilon_k
=\delta\epsilon_k^+ +\delta\epsilon_k^-$ 
may be decomposed into the sum of contributions even (+) and odd (-) with
respect to the parity operation ${\bf k}\rightarrow -{\bf k}$. It
differs in general from the
contribution from the external potential $U_k^{\rm ext}$ through 
vertex corrections, or more physically, through electronic 
polarization potentials \cite{PN}. This fact may be expressed through the 
following diagonal self--consistency relation,
\begin{equation} 
\delta\epsilon_k=U_k^{\rm ext}+2\sum_{\bf p} (V_q+f_{k p}) \delta n_p.
\end{equation}
Here, $V_q=4\pi e^2/q^2$ is the Fourier transform of the long--range Coulomb
interaction and $f_{k p}$ denotes the short--range Fermi--liquid interaction
the consequences of which are, however, not considered in what follows
since we are interested 
in the limit of not too large ${\bf q}$ where the polarization correction from
the long--range Coulomb interaction dominates the diagonal energy change. 

The kinetic equation for the linearized off--diagonal distribution function
$\delta g_k({\bf q},\omega)=g_k({\bf q},\omega)-g_k^0$ reads
\begin{equation}
\delta g_k+\theta_k\delta\Delta_k+\frac{\omega^2-\eta_k^2}{4|\Delta_k|^2}
\lambda_k\delta\Delta_k-\lambda_k\delta\Delta^{\dagger}
\frac{\Delta_k}{|\Delta_k|}=\frac{\lambda_k}{2|\Delta_k|}
\left[\omega\delta\epsilon_k^+ +\eta_k\delta\epsilon_k^-\right].
\end{equation}
A straight--forward variation of the equilibrium gap equation (6) leads to 
an off--diagonal self--consistency relation
\begin{equation}
\delta\Delta_k({\bf q},\omega)=\sum_{\bf p} V_{k p}\ 
\delta g_p({\bf q},\omega) \nonumber ,
\end{equation}
which can now be used to compute the {\it off--diagonal} energy shifts , 
namely the order parameter
fluctuations $\delta\Delta_k({\bf q},\omega)$ and $\delta\Delta_k^{\dagger}
(-{\bf q},-\omega)$. They  represent the collective oscillations necessary to
maintain gauge invariance, and must be determined self consistently with the
off--diagonal kinetic equation \cite{B-MN,wolfle} (from which
particle--hole asymmetric terms, which are typically of the order 
$O(T/T_{\rm F})$, with $T_{\rm F}$ the Fermi temperature, have been omitted),
\begin{equation}
\sum_{\bf p} V_{kp}\lambda_p
\frac{\omega^2-\eta_{p}^2}{4|\Delta_p|^2}\delta\Delta_p^-
=\sum_{\bf p} V_{kp}\lambda_p
\frac{\omega\delta\epsilon_p^+ +\eta_p\delta\epsilon_p^-}{2|\Delta_p|}.
\end{equation}
In deriving (15) the equilibrium gap equation (6) has been used for
simplification. 
In case of particle--hole symmetry, the density, current and Raman
fluctuations do not couple to the quantity $\delta\Delta^+$,
which represents the amplitude fluctuations of the order parameter.
The physical significance of the quantity $\delta\Delta_k^-$ becomes clear
in the (macroscopic) limit $\omega\ll 2\Delta_0$, in which only fluctuations of
the phase $\phi$ of the superconducting  energy gap determine the dynamics of
the condensate and $\delta\Delta_k^-({\bf q},\omega) =|\Delta_k|
i\delta\phi({\bf q},\omega)$. 

The important macroscopic observables, namely the density 
fluctuations $\delta n_1({\bf q},\omega)$, charge fluctuations 
$\delta n_e({\bf q},\omega)$,  the Raman fluctuations $\delta
n_{\gamma}({\bf q},\omega)$ and the current fluctuations
${\bf j}({\bf q},\omega)$ are defined as 
\begin{eqnarray}
\delta n_1({\bf q},\omega) &=& 2 \sum_{\bf k} 1 \ 
\delta n_k({\bf q},\omega), \nonumber \\
\delta n_e({\bf q},\omega) &=& 2 \sum_{\bf k} e \ 
\delta n_k({\bf q},\omega), \nonumber \\
\delta n_{\gamma}({\bf q},\omega) &=& 2 \sum_{\bf k} \gamma_k\delta 
n_k({\bf q},\omega), \\ 
{\bf j}({\bf q},\omega)&=&2\sum_{\bf k}{\bf v}_k[\delta n_k({\bf q},\omega)
+\phi_k\delta\epsilon_k({\bf q},\omega)]. \nonumber
\end{eqnarray}
The factors of 2 arise from spin degeneracy. 

We proceed with a solution of Eq. (15). For the time being we would like to
restrict ourselves to the case where the pairing interaction factorizes as 
\begin{equation} 
V_{k p}=-V \frac{\Delta_k\Delta_p}{\Delta_0^2}.
\end{equation}
This ansatz is sufficiently general to allow for an equilibrium gap 
function $\Delta_k$ of arbitrary anisotropy in ${\bf k}$--space. The 
maximum of such a gap is denoted $\Delta_0$. $\Delta_k$ 
is determined from the following form for the equilibrium gap equation 
(see Appendix A for further details),
\begin{equation} 
\frac{1}{V}=\sum_{p\in {\rm shell}}\theta_p\frac{|\Delta_p|^2}{\Delta_0^2}.
\end{equation}
Using Eq. (17), Eq. (15) can be solved immediately to give
\begin{equation}
\frac{\delta\Delta^-({\bf q},\omega)}{2|\Delta_k|}
=\frac{i}{2}\delta\phi({\bf q},\omega)=
\frac{\sum_{\bf p}\lambda_p(\omega\delta\epsilon_p^++\eta_p\delta\epsilon_p^-)}
{\sum_{\bf p}\lambda_p(\omega^2-\eta_p^2)}.
\end{equation}
The physical significance of the result (19) is that it describes the
Goldstone mode for superconductors, the Anderson--Bogoliubov or
gauge mode, i. e. the massless collective mode related to the spontaneously
broken gauge symmetry. It is the existence of this mode which guarantees
gauge invariance of the response theory or, equivalently, charge
conservation, which we would like to briefly demonstrate now. For this
purpose let us write the kinetic equation (8) in a form in which the l.h.s
is reminiscent of the usual Landau--Silin equation \cite{PN},
\begin{equation}
\omega\delta n_k-{\bf q}\cdot{\bf v}_k[\delta n_k
-\frac{\partial n_k^0}{\partial \xi_k}\delta\epsilon_k]=
-\lambda_k[\omega\delta\epsilon_k^++\eta_k\delta\epsilon_k^-]
+\lambda_k[\omega^2-\eta_k^2]\frac{\delta\Delta_k^-}{2|\Delta_k|}.
\end{equation}
Inserting Eq. (19), one observes that the r.h.s. of Eq. (20) vanishes upon
summation over momenta ${\bf k}$ and the l.h.s. of (20) generates the
continuity equation for the particle number density, 
\begin{equation} 
\omega\delta n_1-{\bf q}\cdot{\bf j}=0.
\end{equation}
Hence we have demonstrated that accounting properly for the fluctuations
of the phase of the order parameter leads to the conservation law for the
(charge) density. 
Alternatively, one could argue in the following way: 
the phase fluctuation term $\delta\Delta_k^-$ on the r.h.s. of (20) can be
thought of having its origin in a replacement of the scalar ($\Phi^{\rm
ext}$) and vector (${\bf A}^{\rm ext}$) potentials in the first term on the 
r.h.s. of Eq. (20), representing the external 
potential contributions to $\delta\epsilon_k^+$ and $\delta\epsilon_k^-$,
by their gauge--invariant counterparts 
\begin{eqnarray}
\Phi^{\rm ext}&\rightarrow&\Phi^{\rm ext}-\frac{1}{c}
\frac{\partial \chi}{\partial t}, \nonumber \\
{\bf A}^{\rm ext}&\rightarrow&{\bf A}^{\rm ext}+\nabla\chi \ \ \ .
\end{eqnarray}
The phase variable $\chi$ characterizing this gauge transformation can then
be fixed by the requirement of charge conservation, which,
together with the trivial connection between $\chi$ and the order
parameter phase fluctuation $\delta\phi$,  
$$
\delta\phi=-\frac{2e}{  c}\chi ,
$$
immediately leads to the result (19). This demonstrates clearly the
equivalence of the existence of the Anderson--Bogoliubov mode with
gauge invariance as well as the connection of gauge--invariance with the
conservation law for the (charge) density. 

Next we would like to demonstrate two simple consequences of a 
gauge--invariant formulation of superconducting response theory, namely 
those for the
current response in the static ($\omega\rightarrow 0$) and for the Raman
response in the homogeneous (${\bf q}\rightarrow 0$) limit. 
The static limit of the kinetic equation (20) reads
\begin{equation}
\delta n_k - \frac{\partial n_k^0}{\partial\xi_k}\ \delta\epsilon_k=
\lambda_k\delta\epsilon_k^--\lambda_k\eta_k
\frac{\sum_{\bf p} \lambda_p\eta_p\delta\epsilon_p^-}{\sum_{\bf p} \lambda_p\eta_p^2}.
\end{equation}
Integrating this according to the prescription (16) one gets the
static (super) current response expressed in a generalized London formula:
\begin{eqnarray}
{\bf j}_e({\bf q},0)=e{\bf j}({\bf q},0)&=&
-\frac{e^2}{c}\stackrel{\leftrightarrow}{\Delta}({\bf q},0){\bf A}^{\rm ext}
({\bf q},0),\nonumber \\
\stackrel{\leftrightarrow}{\Delta}({\bf q},\omega)&=&
\left\{\stackrel{\leftrightarrow}{\lambda}-
\frac{\stackrel{\leftrightarrow}{\lambda}\cdot{\bf q}:
{\bf q}\cdot\stackrel{\leftrightarrow}{\lambda}}{{\bf q}
\cdot\stackrel{\leftrightarrow}{\lambda}\cdot{\bf q}}\right\}
({\bf q},\omega), \\
\stackrel{\leftrightarrow}{\lambda}({\bf q},\omega)&=&2\sum_{\bf p} 
\lambda_p \ {\bf v}_p : {\bf v}_p , \nonumber
\end{eqnarray}
Clearly, the second term in (24), sometimes referred to as backflow term 
\cite{PJHDE}, is necessary to maintain charge conservation 
in the general case of an anisotropic superconductor. Equivalently, it
guarantees that the current is purely transverse in the static limit. 

Let us now turn to the Raman response in the homogeneous limit. Here we
may ignore terms linear in $\Phi^{\rm ext}$ and ${\bf A}^{\rm ext}$. For
${\bf q}\rightarrow 0$ the kinetic equation (20) assumes then the strikingly
simple form
\begin{equation}
\delta n_k=-\lambda_k\delta\epsilon_k^+ +
\frac{\sum_{\bf p} \lambda_p\delta\epsilon_p^+}{\sum_{\bf p} \lambda_p},
\end{equation}
in which the second term on the r.h.s. originates from the order parameter
phase fluctuations. The Coulomb interaction becomes irrelevant in this
limit, as will become clear later, and we may write
$\delta\epsilon_k^+=\gamma_k\delta\epsilon_{\gamma}$.
The homogeneous Raman response of anisotropic superconductors can then be
written in a form analogous to the London limit of the current response as
\begin{eqnarray}
\delta n_{\gamma}(0,\omega)&=&\Delta_{\gamma\gamma}(0,\omega)
\delta\epsilon_{\gamma}(0,\omega), \nonumber \\
\Delta_{a b}({\bf q},\omega)&=&
-\left\{\lambda_{a b} 
-\frac{\lambda_{a 1}\lambda_{1 b}}{\lambda_{1 1}}\right\}({\bf q},\omega), \\
\lambda_{a b}({\bf q},\omega)&=&2\sum_{\bf p} \lambda_p \ a_p b_p  
\ \ \ ;  \ \ \ a_p,b_p=1,\gamma_p. \nonumber
\end{eqnarray}
As in the case of the current response there is a ``backflow'' term 
(the second in the curly brackets), which guarantees charge conservation. 
This is easily seen in the limit of parabolic bands, where the Raman
vertex $\gamma_k$ is a constant and as a consequence the two terms
in curly brackets cancel precisely. This is just another way of stating
that there are no density fluctuations possible in the
homogeneous limit ${\bf q}\rightarrow 0$ of a superconductor. 

It is worth noting that there exists a homogeneous limit of the 
electronic Raman response also in normal metals when quasiparticle scattering
processes are important characterized by a momentum--dependent lifetime
$\tau_k$. In Appendix B we show that an equation similar to (26) holds
in the normal state with the Tsuneto function $\lambda_k$ replaced by
$\Lambda_k=(-\partial n_k^0/\partial\xi_k)(1-i\omega\tau_k)^{-1}$.

Let us now turn to a description of the Raman response at finite
wavevector ${\bf q}$. Our starting point will be Eq. (8) in which we select
the even--parity contributions $\delta n_k^+$ and
$\delta\epsilon_k^+=\delta\epsilon_0+\gamma_k\delta\epsilon_{\gamma}$
with $\delta\epsilon_0=V_q\delta n_1$:
\begin{eqnarray}
\delta n_k^+&=&[S_k+\Omega^2\lambda_k]\delta\epsilon_0
+[\gamma_k S_k+\Omega^2\lambda_k
\frac{\lambda_{\gamma 1}}{\lambda_{11}}]\delta\epsilon_{\gamma}, \nonumber \\
S_k&=&\frac{\eta_k^2\phi_k-\omega^2\lambda_k}{\omega^2-\eta_k^2}, \\
\Omega^2&=&\frac{\omega^2}{\omega^2-\omega_q^2} 
\ \ \ ; \ \ \  \omega_q^2=\frac{{\bf q}\cdot\stackrel{\leftrightarrow}{\lambda}\cdot{\bf q}}{\lambda_{1 1}}. \nonumber
\end{eqnarray}
Integration over momenta ${\bf k}$ in Eq. (27) yields equations that describe
the coupling of density and Raman response:
\begin{eqnarray}
\delta n_1&=&\chi_{1 1}^{(0)}\delta\epsilon_0
+\chi_{1 \gamma}^{(0)}\delta\epsilon_{\gamma}, \nonumber \\
\delta n_{\gamma}&=&\chi_{\gamma 1}^{(0)}\delta\epsilon_0
+\chi_{\gamma \gamma}^{(0)}\delta\epsilon_{\gamma}, 
\end{eqnarray}
where we have defined generalized (free) response functions
\begin{equation}
\chi_{a b}^{(0)}=\Delta_{a b}+
\frac{\omega_q^2}{\omega^2-\omega_q^2}
\frac{\lambda_{a 1}\lambda_{1 b}}{\lambda_{1 1}}+
2\sum_{\bf p}\left\{\frac{\eta_p^2(\phi_p-\lambda_p)}{\omega^2-\eta_p^2}a_p b_p\right\}.
\end{equation}
The quantities $\chi_{a b}^{(0)}({\bf q},\omega)$ are straightforward
generalizations of the free superconducting Lindhard function which include 
vertices $a_k$, $b_k$. The Anderson--Bogoliubov
collective mode causes the second term in the transverse contribution
$\Delta_{a b}$ 
(note that $\Delta_{a 1}$ =$\Delta_{1 b}$= $\Delta_{1 1} \equiv 0$)
and the longitudinal term which is characterized by the
gauge mode frequency $\omega_q$. 
Finally we explicitely work out the Coulomb renormalization
$\delta\epsilon_0=V_q\delta n_1$ which couples Raman and density
fluctuations and arrive at the following final result:
\begin{eqnarray}
\delta n_{\gamma}&=&\chi_{\gamma\gamma}\delta\epsilon_{\gamma}, \nonumber \\
\chi_{a b}&=&\chi_{a b}^{(0)}
-\frac{\chi_{a 1}^{(0)}\chi_{1 b}^{(0)}}{\chi_{1 1}^{(0)}}
+\frac{\chi_{a 1}^{(0)}\chi_{1 b}^{(0)}}{\chi_{1 1}^{(0)}}
\frac{1}{\epsilon}, \\
\epsilon&=&1-V_q\chi_{1 1}^{(0)}. \nonumber
\end{eqnarray}
It should be noted that the Raman response in superconductors in the 
small $q$ limit in a form equivalent to Eq. (30) has first been derived by
Klein and Dierker \cite{klein} using the Greens--function method. Our result 
from the kinetic equation method looks slightly different and, particularly,
one can see that the Coulomb interaction acts so as
to split the Raman response into an unscreened ``transverse'' and a
dielectrically screened ``longitudinal'' part, the latter being described
by the full dielectric function of $\epsilon({\bf q},\omega)$ of the
superconductor in complete analogy to the behavior of the current response,
discussed in Ref. \cite{PJHDE}. 
Furthermore, an inspection of Eq. (30) shows, that all
terms in the full response function 
$\chi_{a b}$ except $\Delta_{\gamma\gamma}$ are at least of
the order $O(q^2)$, the last (longitudinal) term being even smaller, of
the order $O(q^2/\epsilon)$. The role of the Coulomb interaction is thus
limited to show up in terms of the order of at most $O(q^2)$ and is seen to be
negligible in the homogeneous limit. The role of the collective
Anderson--Bogoliubov mode, on the other hand, mainly consists of providing a
particle number conservation law. 
This manifests itself first in the correct mass fluctuation limit
of Eq. (30), in which $\gamma_k$=const and all transverse terms vanish
identically leaving essentially the screened RPA Lindhard response of the
superconductor in which the collective mode gets shifted to the plasma
frequency. Second it manifests itself in ``partial screening'' effects in
the homogeneous limit of the Raman response, described by the second term of
$\Delta_{\gamma\gamma}$. These effects will be discussed in detail later. 

\subsection{Final Results for ${\bf q} \rightarrow 0$.}

Since $q\xi << 1$ in the cuprates (with $\xi$ the coherence length), we are 
mostly interested in Raman scattering with vanishing momentum transfers.
For such a case 
it is essential to conclude at this stage that the most important
contribution to the electronic Raman effect in superconductors in the
small--${\bf q}$ collisionless limit comes from the response function
$\Delta_{\gamma\gamma}(0,\omega)=\lim_{{\bf q}\rightarrow 0}\chi_{\gamma\gamma}
({\bf q},\omega)$. 
Physically, this function describes 
the photon--induced breaking of a Cooper pair into a pair of Bogoliubov
quasiparticles with total momentum ${\bf q}$ which is approximately zero.

Writing the Raman vertex $\gamma$ as a sum
$\gamma({\bf k})=\gamma_{0}+\delta\gamma({\bf k})$
of an isotropic and an anisotropic part (using Fermi surface harmonics
$\delta\gamma({\bf k})=\sum_{L \ne 0}\gamma_{L}({\bf k}))$, 
one may decompose the screened Raman
response function at zero temperature in the limit of small
momentum transfers as $\chi_{\gamma\gamma}=\chi_{\parallel}+\chi_{\perp}$ with
\begin{eqnarray}
\chi_{\parallel}&=&
\frac{[\chi_{\gamma 1}^{(0)}]^2}{\chi_{1 1}^{(0)}}
\frac{1}{\epsilon}=O\left(\frac{q^{2}}{\epsilon}\right), \nonumber \\
\chi_{\perp}&=&\Delta_{\delta\gamma\delta\gamma}(0,\omega)+O(q^{2})
\end{eqnarray}
into a longitudinal part ($\parallel$), affected by
(longitudinal) screening through the
dielectric function $\epsilon$ of the superconductor, and a
transverse part ($\perp$), independent of $\epsilon$. Thus for
${\bf q} \rightarrow 0$, only the transverse part of $\chi$ remains.
We also see that 
in the case of the isotropic density
vertex, $\gamma({\bf k})=\gamma_{0}$,
the long range Coulomb forces completely screen the response
and thus the only contribution to Raman scattering at $q=0$
comes from energy bands with nonparabolic dispersion, i.e.,
the $L=2$ and higher terms, representing intracell charge
fluctuations. 

Taking the limit of ${\bf q} \rightarrow 0$ and carrying out the $\xi$
integration in the Tsuneto function, we obtain the final result for 
the Raman response at finite temperatures,
\begin{equation} 
\chi({\bf q} \rightarrow 0,i\omega)=\chi_{\delta\gamma,\delta\gamma}(i\omega)
-{\chi_{\delta\gamma,1}^2(i\omega)\over{\chi_{1,1}(i\omega)}},
\end{equation}
where the spectrum of $\chi_{a,b}$ is given by
\begin{equation}
\chi_{a,b}^{\prime\prime}(\omega)={\pi N_{F}\over{\omega}}
\tanh\left(\frac{ \omega}{4T}\right)
Re\left< {a({\bf k})b({\bf k})\mid\Delta({\bf k})\mid^{2}\over{\sqrt{
\omega^{2}-4\mid\Delta({\bf k})\mid^{2}}}}\right>
\end{equation}
where $N_{F}$ is the density of states for both spin projections, $Re$
denotes taking the real part, and
$\langle \cdots \rangle$ denotes performing an average over the
Fermi surface defined by $$\langle A({\bf k}) \rangle={\int d^{2}k \delta(E_{F}
-\epsilon({\bf k})) A({\bf k})\over{\int d^{2}k \delta(E_{F}
-\epsilon({\bf k}))}}.$$
The real part of $\chi$ is given as
\begin{equation}
\chi_{a,b}^{\prime}(\omega)=\langle a({\bf k})b({\bf k}) A^{\prime}
(\omega) \rangle \tanh\left(\frac{ \omega}{4T}\right),
\end{equation}
$$
A^{\prime}(\omega)=N_{F}{\mid\Delta({\bf k})\mid^{2}\over{\omega}}
\cases{ {2\over{\sqrt{\mid\Delta({\bf k})\mid^{2}-(\omega/2)^{2}}}}
\arctan\left(\omega\over{2\sqrt{\mid\Delta({\bf k})\mid^{2}-
(\omega/2)^{2}}}\right), &$\mid\Delta({\bf k})\mid^{2} > (\omega/2)^{2},$\cr
{1\over{\sqrt{(\omega/2)^{2}-\mid\Delta({\bf k})\mid^{2}}}}
\log\left[{\omega/2-\sqrt{(\omega/2)^{2}-\mid\Delta({\bf k})\mid^{2}}\over{
\omega/2+\sqrt{(\omega/2)^{2}-\mid\Delta({\bf k})\mid^{2}}}}\right],
&$\mid\Delta({\bf k})\mid^{2} \le (\omega/2)^{2}$.} 
$$
This is the expression for the gauge invariant Raman response which has 
Coulomb screening and the Anderson--Bogoliubov gauge mode
taken into account.  It is also the form for the response calculated 
diagrammatically in 
the ``pair approximation'' for the bare bubble, modified with the usual
R.P.A. screening terms \cite{klein,ag}. It does not take into account any
vertex corrections resulting from the pairing interaction in other channels
other than the pairing channel, as explained in the preceeding
discussion. We will refer back to this expression in the following Sections.

        Important information can also be obtained from the temperature
dependence of the response in the limit of zero
frequency shifts, i.e., the static reponse.  It can be shown that the
ratio of the response in a superconductor to that of a normal metal in
the limit of vanishing frequencies is given by the simple expression
\begin{equation}
{\chi^{\prime\prime}_{s.c.}(\omega \rightarrow
0)\over{\chi^{\prime\prime}_{n.s.}(\omega \rightarrow 0)}} = {2\langle
f(\mid\Delta({\bf  \hat k})\mid) \mid \delta\gamma({\bf  \hat k})
\mid^{2}\rangle\over{\langle \mid \delta\gamma({\bf  \hat
k})\mid^{2}\rangle}},
\end{equation}
where $f$ is a Fermi function \cite{me}. 
This is an exact result which does not depend on vertex
corrections and only depends appreciably
on impurity scattering for nearly resonant 
impurity scatterers \cite{me,imp}.

The important feature in all these expressions is that in
general a coupling of the Raman vertex to the energy gap can
occur under the  momentum averaging over the Fermi surface. In
all previous studies, this $k$--dependence was either ignored or
not fully exploited to determine important information
concerning the symmetry of the energy gap.
This will be explicitly demonstrated in the following section 
where we evaluate these expressions for
the screened Raman response for various pair state candidates
and discuss its relevance to the cuprate materials. 

We close this section by returning to the question of the presence
of massive modes and final--state vertex corrections (see above). 
In principle, the massive modes can lie at low
frequencies and affect the low frequency behavior of correlation
functions, and in particular, could even be used as a signal for
a certain type of order parameter symmetry.
We have carried out an analysis of the full gauge invariant
calculation in the Appendix C for both cylindrical and spherical 
Fermi surfaces for a generalized pairing interaction. 
We identified the position of the
collective modes as a function of coupling strength for a gap of
$d_{x^{2}-y^{2}}$ ($\Gamma_{3}^{+}$ representation using the
notation of Sigrist and Rice, \cite{sigrist})
symmetry for both a spherical and cylindrical
Fermi surface.  Our conclusions are threefold: 1) the Goldstone modes
do affect the Raman spectrum in the limit of ${\bf q}\rightarrow 0$, 
in certain polarization symmetries ($A_{1g}$) 
where the "backflow term" in (32) is finite, 2)
optical (massive) collective modes couple to a light probe only
for case of a Fermi surface with $z$--dispersion for $d$--wave
vertex corrections in the $B_{1g}$ and $E_{g}$ channels, 3) these
modes are significantly damped and have a vanishing residue for
those modes which lie at low energies. Therefore, the collective
modes are of little importance to the Raman spectrum. These
conclusions can be generalized to other energy gaps which have
line nodes on the Fermi surface. Also, we discuss the role of the
final--state interactions and find that while the overall shape of the
spectra can be affected, the corrections are relatively minor.
The details are contained in Appendix C, including a more general
discussion of the role of vertex corrections. Therefore, we can conclude
that Eqs. (32-35) give an adequate desciption of the Raman response.

\section{Band Structure, Fermi Surface and the Raman Vertex for Tetragonal
Symmetry}

In this section we aim at providing a link between the Raman vertex and
band structure for tetragonal crystals. In particular, we show how the
choice of light polarization orientations results in selection rules for
the symmetry components of the Raman tensor. In the first subsection we
only consider scattering within a single band while we consider the case
of multiple bands at the Fermi surface in the following subsection. 

\subsection{Single Band}
As we have seen in the
previous section (see Eq. (2)), the Raman tensor is directly related to
the curvature of the band dispersion. In the following we limit our
considerations to tetragonal materials which are relevant to high 
T$_c$ superconductors. Although these materials possess
orthorhombic distortions away from tetragonality, the selection
rules for example from phonon scattering in the normal state
seem to indicate that these orthorhombic distortions are small
and that a tetragonal symmetry classification can be employed with
little loss of generality. A simplest choice for the band structure which
contains the basic physics of 2D--like tetragonal systems with lattice
constant $a$ is given by,
\begin{equation} %
\epsilon({\bf  k})=-2t\left[\cos(k_xa)+\cos(k_ya)\right]
                  +4t^{\prime}\cos(k_xa)\cos(k_ya).
\end{equation}
Here $t$ and $t^{\prime}$ are the nearest and next nearest neighbor
hopping parameters, respectively. This is the anti--bonding band derived
from a reduction of a three band model \cite{fred}, 
which gives by far
the largest contribution to the density of states at the Fermi level for
the cuprate systems \cite{pickett}.
The 2D Fermi surface is defined through
the relation $\epsilon({\bf  k})=\mu$, where $\mu$ is the chemical
potential, which in turn determines the Fermi momentum,
 \begin{equation}  %
{\bf  k}_F=k_F(\varphi)\left(
\begin{array}{c}
\cos(\varphi)\\ \sin(\varphi)
\end{array}\right).
\end{equation}
The scalar prefactor $k_F(\varphi)$ can be expanded with respect to the
fully symmetric basis functions ($A_{1g}$ or $\Gamma_{1}^{+}$ using the
notation of Sigrist and Rice \cite{sigrist})
for the tetragonal $D^{4h}$ point group, 
\begin{eqnarray}
k_F(\varphi)&=&k_F^{(0)}+k_F^{(2)}\cos(4\varphi)+
                         k_F^{(4)}\cos(8\varphi)+\cdots \nonumber \\
            &=&k_F^{(0)}+\sum_{L=2,{\rm even}}k_F^{(L)}a_L^0(\varphi).
\end{eqnarray}
The higher order Fourier coefficients $k_F^{(L)}$ for $L>0$
take into account deviations from cylindricity of the Fermi surface,
$k_F^{(0)}$.  The basis functions for the irreducible
representations of the point group symmetry can be generalized as
\begin{equation}
a_L^j(\varphi)=\cos\left(2L\varphi+ \frac{j\pi}{2}\right), \ \ \
                                        j\in \{0,1\},
\end{equation}
where $a_L^j$ transforms according to the $A_{j+1 g}$--symmetry, and
\begin{equation}
b_L^j(\varphi)=\cos\left[(2L-2)\varphi+ \frac{j\pi}{2}\right], \ \ \
                                        j\in \{0,1\},
\end{equation}
which transform according to the $B_{j+1 g}$--symmetry \cite{tinkham}. 

The Raman tensor is given by
\begin{eqnarray}
&{\stackrel{\leftrightarrow}{\gamma}}({\bf  \hat k})&=
m M^{-1}({\bf  \hat k})\nonumber\\
                  &=&2ma^{2}
\left(\begin{array}{cc}
\cos(k_{Fx}a)[t-2t^{\prime}\cos(k_{Fy}a)]
&2t^{\prime}\sin(k_{Fx}a)\sin(k_{Fy}a) \\
2t^{\prime}\sin(k_{Fx}a)\sin(k_{Fy}a)
& \cos(k_{Fy}a)[t-2t^{\prime}\cos(k_{Fx}a)]
\end{array}\right)
\end{eqnarray}
where $M^{-1}_{\alpha\beta}=
\partial\epsilon({\bf k})/\partial k_{\alpha}\partial k_{\beta}$.
We denote 
matrices in $k-$space (except the Pauli matrices)
with a double--arrow.
We proceed to expand the Raman tensor in quarternions,
\begin{equation}
{\stackrel{\leftrightarrow}{\gamma}}({\bf  \hat k})=m{\stackrel{\leftrightarrow}{M}}^{-1}({\bf  \hat k})=
\sum_{m=0}^{3}\delta^{(m)}({\bf \hat k})\ \tau^m,
\end{equation}
with $\tau^m$ 
the Pauli matrices in $2$--D\ ${\bf k}$--space. 
The polarization light vectors
select elements of the Raman tensor according to
\begin{equation}
\gamma({\bf  \hat k})={\hat e}^I{\stackrel{\leftrightarrow}{\gamma}}
({\bf \hat k}){\hat e}^S 
=\sum_m({\hat e}^I \tau^m {\hat e}^S)\ \delta^m({\bf \hat k}).
\end{equation}
Next we expand the functions $\delta^{(m)}({\bf \hat k})$ in terms of 
Fermi surface harmonics
\begin{equation}
\delta^{(m)}({\bf  \hat k})=\sum_{L=2,{\rm even}}
\delta_L^m\phi_L^m(\varphi),
\end{equation}
where
\begin{equation}
\phi_L^m(\varphi)=\cases{
a_L^0(\varphi), &$A_{1g} \ (m=0),$\cr
b_L^1(\varphi), &$B_{2g} \ (m=1),$\cr
a_L^1(\varphi), &$A_{2g} \ (m=2),$\cr
b_L^0(\varphi), &$B_{1g} \ (m=3)$.}
\end{equation}
The coefficients ${\hat e}^{\rm I} \tau^m {\hat e}^{\rm S}$ of the Raman vertex
can only assume the values 0 or 1 and have the following explicit form:
\begin{equation}
{\hat e}^{\rm I} \tau^m {\hat e}^{\rm S}=\cases{
[e_x^Ie_x^S+e_y^Ie_y^S], &$A_{1g} \ (m=0),$\cr
[e_x^Ie_y^S+e_y^Ie_x^S], &$B_{2g} \ (m=1),$\cr
[e_x^Ie_y^S-e_y^Ie_x^S], &$A_{2g} \ (m=2),$\cr
[e_x^Ie_x^S-e_y^Ie_y^S], &$B_{1g} \ (m=3)$.}
\end{equation}
We now can reconstruct the Raman vertex (see Eq. (2)) as
\begin{eqnarray}
\gamma({\bf  \hat k})&=&\sum_{L,m}\gamma_L^m \ \phi_L^m(\varphi) \nonumber \\
\gamma_L^m&=&{\hat e}^I \tau^m {\hat e}^S \ \ \delta_L^{(m)}.
\end{eqnarray}
We therefore see the connection between polarization orientation and
symmetry. For example, if ${\hat e}^I={\hat e}^S=\hat x$ are chosen, then
a combination of $A_{1g}$ and $B_{1g}$ is selected. If
${\hat e}^I={\hat e}^S=2^{-1/2}(\hat x+\hat y)$ then $A_{1g}$ and $B_{2g}$
symmetries are combined. These examples show in particular that $A_{1g}$
symmetry cannot be individually accessed for in--plane polarizations, and
subtraction procedures must be used.

\subsection{Multi--sheeted Fermi surfaces}
The above consideration can be easily adopted to the case of multiple
energy bands contributing to the Fermi surface, for instance, 
due to the contribution of the chains as well as the planes
in YBCO\cite{pickett}. The total electronic Raman cross section for intraband 
scattering is just the sum of the contribution to the cross section for
each band and thus results from the addition of the Raman response functions,
\begin{equation}
\chi_{total}=\sum_{\alpha}\chi_{\alpha}\{\Delta_{\alpha}\},
\end{equation}
where $\chi_{\alpha}\{\Delta_{\alpha}\}$ denotes the contribution to the
scattering from band $\alpha$ with an energy gap $\Delta_{\alpha}$ for
each band. The Raman vertex for each band is separately calculable in
the same manner as in the previous section, and the same considerations
there can be carried over to the multi--sheeted case trivially. 

We draw
the important point however that the contribution from each band will be
weighted by the relative density of states and curvature of each band at
the Fermi level. 
Therefore, for
the case of the cuprates we believe that by far the largest contribution
to electronic Raman scattering is thus arising from the single anti--bonding
Cu-O layer band with the largest density of states and greatest curvature
at the Fermi level. Further, 
it is not even
clear whether an energy gap exists on the other energy bands (e.g., 
associated with the chains in YBCO), and if it does exist what symmetry
it has. Therefore, we feel that to a very good approximation, the 
intraband scattering has a dominant contribution from the anti--bonding
band. This is certainly the case for the single Cu-O layer compound,
Tl$_{2}$Ba$_{2}$CuO$_{6}$.

We close this subsection with a discussion concerning interband scattering,
or scattering from the anti--bonding band to another band near or at
the Fermi level. In this case a separate theory is necessary to take
into account the separate band dispersion and energy difference of the
two bands, and the energy gap on each of the bands. This at present remains
a future consideration. However, we remark at this point that once again
this contribution to the scattering cross section will be smaller than
the contribution from scattering within the anti--bonding band due to the
small curvature of the other energy bands and the value of the density 
of states at the Fermi level (again, the chains for example).
Also, since the nature
of the energy gap on the chains is completely unclear, we consider 
interband scattering to be beyond the scope of the present manuscript
\cite{bands}. 

\section{Predictions for various pair states}
In this Section we evaluate Eqs. (32) and (33) for various pair states
that are discussed in the literature as candidates to describe
the pairing symmmetry in the cuprates. We will first discuss the
simple case of an isotropic gap and the discuss $d$--wave gaps
(in particular, we will focus on the $d_{x^{2}-y^{2}}$ pairing
symmetry) and then discuss gaps which are anisotropic but do not
contain line nodes, e.g., an $s+id$ and an anisotropic $s$--wave gap.

\subsection{Isotropic $s$--wave}

For the case of an angular independent gap,
$\Delta({\bf k})=\Delta_{0}$, the Raman response is given by the
simple BCS expression. It has the stereotypic features of BCS
theory, namely, the existence of an energy gap threshold
$2\Delta$ required to break a Cooper pair, and the ubiquitous
square root divergence associated with the gap edge. We note in
particular that the Raman vertex couples trivially to the energy
gap and just determines an overall prefactor governing the Raman
intensity. The vertex does not affect the lineshape and thus the spectrum
is polarization independent apart from an overall prefactor. 
A polarization dependence can be
generated in BCS theory by taking into account channel--dependent
final--state interactions \cite{klein} and/or impurity
scattering \cite{me}, and accurate fits to the Raman data on A-15
superconductors can be obtained. However, for the most part this only
produces an channel dependence in the vicinity of the gap edge
and thus the main feature of the response is the uniform gap
existing for all polarizations, which clearly cannot give an
adequate description to the Raman spectra of the cuprate materials.

\subsection{$d_{x^{2}-y^{2}}$ pairing}

    The explicit symmetry dependence of the spectra results due
to a coupling of the energy gap and vertex when the energy gap
is anisotropic. To demonstate this, we now carry
out the averages over a cylindrical Fermi surface in Eq. (33) using a
$d_{x^{2}-y^{2}}$ gap, 
$\Delta({\bf  \hat k},T)=\Delta_{0}(T)\cos(2\varphi)$ (see Appendix A).
We find that the Raman spectrum can be written down 
analytically in terms of complete
elliptical integrals. Taking screening into account and defining 
$x= \omega/2\Delta_{0}$, we obtain for $T=0$,
\begin{eqnarray}
& &\chi_{B_{1g}}^{\prime\prime \ sc.}=
\chi_{B_{1g}}^{\prime\prime}({\bf  q}=0,\omega)= \\
& &{2N_{F}\gamma_{B_{1g}}^{2}\over{3\pi x}}
\times\cases{[ (2+x^{2})K(x)-2(1+x^{2})E(x)], & $x\le 1$
, \cr
x[ (1+2x^{2})K(1/x)-2(1+x^{2})E(1/x)], 
& $x > 1$,}\nonumber
\end{eqnarray}
\begin{eqnarray}
& &\chi_{B_{2g}}^{\prime\prime \ sc.}=
\chi_{B_{2g}}^{\prime\prime}({\bf  q}=0,\omega)= \\
& &{2N_{F}\gamma_{B_{2g}}^{2}\over{3\pi x}}
\times\cases{[ (1-x^{2})K(x)-(1-2x^{2})E(x)], & $x\le 1$
, \cr
x[ (2-2x^{2})K(1/x)-(1-2x^{2})E(1/x)], 
& $x > 1$,}\nonumber
\end{eqnarray}
i.e., the $B_{1g}$ and $B_{2g}$ channels are not affected by 
Coulomb screening. This is consistent with the mass fluctuations being 
only intracell in nature for these symmetry channels.
However, the $A_{1g}$ channel which contains both inter--and intracell
fluctuations is partially screened and is determined via
\begin{equation}
\chi_{A_{1g}}^{sc.}(i\omega)=\chi_{A_{1g},A_{1g}}(i\omega)-
\frac{\chi_{A_{1g},1}^2(i\omega)}{\chi_{1,1}(i\omega)},
\end{equation}
with the spectral functions
\begin{eqnarray}
& &\chi_{A_{1g},A_{1g}}^{\prime\prime}({\bf  q}=0,\omega)=
{2N_{F}\gamma_{A_{1g}}^{2}\over{15\pi x}}\\
& &\times\cases{
[ (7-8x^{2}+16x^{4})K(x)-(7-12x^{2}+32x^{4})E(x)], & $x\le 1$, \cr
x[(11-28x^{2}+32x^{4})K(1/x)-(7-12x^{2}+32x^{4})E(1/x)], & $x > 1$,}
\nonumber
\end{eqnarray}
\begin{equation}
\chi_{A_{1g},1}^{\prime\prime}({\bf  q}=0,\omega)=
{\sqrt{2}N_{F}\gamma_{A_{1g}}\over{3\pi x}} \cases{
[ (1+2x^{2})K(x)-(1+4x^{2})E(x)], & $x\le 1$, \cr
x[ (-1+x^{2})K(1/x)-(1+x^{2})E(1/x)], & $x > 1$,}
\end{equation}
and
\begin{equation}
\chi_{1,1}^{\prime\prime}({\bf  q}=0,\omega)=
{N_{F}\over{\pi x}}
\cases{[ K(x)-E(x)],& $x\le 1$ , \cr
x[ K(1/x)-E(1/x)], & $x > 1$.}
\end{equation}
The real parts are obtainable through Kramers--Kronig transformation or
numerically integrating Eq. (34).
The response functions for finite $T$ are obtained simply 
by multiplying Eqs. (49-54) by the factor $\tanh(\omega/4T)$.
The partial screening of the $A_{1g}$ channel 
arises technically from the observation that the square of the energy 
gap enters into the response function in Eq. (33). For the case of 
$d_{x^{2}-y^{2}}$ pairing symmetry,
the energy gap squared contains a term which transforms according to $A_{1g}$
symmetry which leads to a finite overlap with the $A_{1g}$ vertex in Eq. (32).
This corresponds to partial ``transverse screening'' of the $A_{1g}$
channel, and via this mechanism the intercell fluctuations are removed.
Similar considerations hold if the gap were of another $d$--wave
symmetry other than $d_{x^{2}-y^{2}}$.

These functions are plotted in Fig. (1) for the 3 symmetries indicated.
We immediately see that the spectra is extremely polarization dependent,
in contrast to the case of isotropic $s$--wave superconductors which is
dominated by the square root divergence at the threshold in each channel.
We see that the peak in the Raman spectra
lies at different frequencies
$\omega_{peak} \sim 2\Delta_{0}(T), 1.6\Delta_{0}(T)$ and $1.2\Delta_{0}(T)$
for the $B_{1g}, B_{2g}$ and $A_{1g}$ channels, respectively \cite{better}.  
The symmetry dependence of the spectra is a direct consequence of the 
angular averaging which couples the gap and Raman 
vertex, and leads to constructive (destructive) interference 
under averaging if the vertex and the gap have the same (different) symmetry.  
Thus it has
been reasoned that the symmetry which shows the highest peak position
gives an unique indication of the predominant symmetry of the gap \cite{us}.
The peak
positions can be mildly affected by including interaction vertex
corrections as discussed in Appendix C, with the net result being a
slight upward shift of the peak location in the $B_{2g}$ and $A_{1g}$
channels. Also, while the presence of $z$--dispersion has little
effect on the $B_{1g}$ (apart from cutting off the logarithmic
divergence) and $B_{2g}$ channels, the $A_{1g}$ peak position can be
changed due to the addition of a terms which has its main contribution
at slightly lower frequencies. This effect is small provided that the
Fermi surface is mostly cylindrical.
We refer the reader to Appendix C for details. 

The symmetry dependence is also manifest in the low frequency behavior,
which can be written as
\begin{eqnarray}
\chi^{\prime\prime}_{B_{1g}}(\omega \rightarrow 0)&=&
3N_{F}\gamma_{B_{1g}}^{2}x^{3}/4+O(x^{5}), \nonumber \\
\chi^{\prime\prime}_{B_{2g}}(\omega \rightarrow 0) &=&
N_{F}\gamma_{B_{2g}}^{2}x/2+O(x^{3}), \nonumber \\
\chi_{A_{1g}}^{\prime\prime}(\omega \rightarrow 0) &=&
N_{F}\gamma_{A_{1g}}^{2}x/2+O(x^{3}),
\end{eqnarray}
i.e., the spectrum rises slower in the $B_{1g}$ channel than the
$A_{1g}$ or $B_{2g}$ channels, which have the same linear rise
with frequency.  The power--laws are insensitive to vertex corrections
and arise solely due to topology arguments.
The appearance of power laws are a signature of an energy gap which 
vanishes on lines (points in 2D) on 
the Fermi surface.  However, the channel dependence of the exponents are 
unique to a $d_{x^{2}-y^{2}}$ pair state. These channel-dependent power laws 
have been observed in the electronic contribution to Raman scattering in
BSCCO \cite{us,hackl1}, YBCO \cite{hackl} and TBCO \cite{hackl3,axel} 
and are strong evidence for a $d$--wave gap of 
this symmetry as opposed to $d_{xy}, d_{xz}$ or  $d_{yz}$ symmetry, 
which also have nodes on lines on the Fermi surface \cite{us,better}.

We now investigate the temperature dependence of the theory and
contrast to that of an isotropic $s$--wave superconductor.
Using a weak coupling expression for the temperature dependence of the
energy gap $(2\Delta_{0}/T_{c}=4.2794)$ (see Appendix A), 
we numerically evaluate Eq. (35) for the temperature dependence
of the normalized static response while taking screening into account.
This function describes how the gap in the Raman spectrum
at low energies opens up with cooling below T$_{c}$.
The results are plotted in Fig. (2) as a function of $T/T_{c}$
for a $d_{x^{2}-y^{2}}$ energy gap compared
to a BCS isotropic gap.  The low temperature behavior is given by a
power--law in $T$ for all channels for the $d$--wave case while the 
ubiquitous exponential dependence in $T$ is seen for all channels in the
$s$--wave case. The power--law behavior for the $d$--wave case is channel
dependent, with exponents identical to those of Eq. (55), in the
sense that $\omega$ can be replaced by $T$.
What is remarkable is that the fall off of the
Fermi function at low temperatures is quite slow in those channels which are
orthogonal to the symmetry of the gap, where the 
$A_{1g}$ and $B_{2g}$
channels show a residual broadening at $T/T_{c}=0.3$ of roughly 20
percent of that of the normal state.  
This was argued in the case of electronic Raman scattering to be
further evidence for an energy gap in the cuprate materials
which has predominantly $B_{1g}$ character, due to the observation that a
gap opens up quickly in the $B_{1g}$ channel compared to 
others channels which have been probed via Raman scattering \cite{us}.

As we have remarked before, since the Raman density response function
only depends on the magnitude of the energy gap, it cannot be
sensitive to the phase of the order parameter and thus cannot be
directly used to determine if the gap changes sign around the
Fermi surface.  While the power--law behavior at low frequencies
and/or temperatures is indicative to the presence of nodes, in
principle a very highly anisotropic order parameter with a small 
uniform gap everywhere on the Fermi surface could mimic the behavior 
of a gap with nodes when inelastic
scattering or experimental resolution smears the threshold. In
principle the detection of a threshold can then only be
performed at very low temperatures where activated behavior can
be observed. Since this remains a possibility 
we now discuss two types of energy gaps which are
anisotropic but have a finite gap around the Fermi surface.

\subsection{$s+id$ pairing}

Kotliar and Joynt have suggested the possibility
that an order parameter which is a superposition of an $s$--wave
and a $d$--wave gap can also provide an adequate description to
the various transport and thermodynamic measurements on the
cuprate systems \cite{sid}. The $s+id$ state has the interesting feature
that in pure tetragonal superconductors there would be two transition
temperatures associated with the formation of each gap separately. This
is a consequence of the admixture of two different representations of
the energy gap. However, orthorhombic distortions remove the $x 
\leftrightarrow y$ symmetry and thus $A_{1g}$ and $B_{1g}$ belong to
the same representation. This leads then to one transition temperature.
Since the orthorhombic distortions are quite small in the cuprates
(judging from the observed phonons and selection rules), the transition
temperature will be broad, which is in some conflict with the 
resistive transitions seen in the cuprates.

Nevertheless, we investigate what such a gap predicts for the
Raman response by evaluating Eq. (33) on a cylindrical Fermi surface
using a gap of the form
$\Delta({\bf \hat k})=\Delta_{s}(T)+i\Delta_{d}(T)\cos(2\varphi)$.
The results can written again in an analytic form in terms of complete
elliptical integrals. Taking screening into account and defining 
$x^2=[( \omega/2)^2-\Delta_s^2]/\Delta_d^2$, 
we obtain for $T=0$,
\begin{eqnarray}
\chi_{B_{1g}}^{\prime\prime \ sc.}=
\chi_{B_{1g}}^{\prime\prime}({\bf  q}=0,\omega)=\Theta(x^{2})
{4N_{F}\Delta_{d}\gamma_{B_{1g}}^{2}\over{3\pi \omega}}\ \ \ \ \ \ \ \ \ \ 
\ \ \ \ \ \ \ \ \ \ \ \ \ \ \ \ \ \ \ \ \ \ \ \ \ \ \ \ \ \ \ \  \\
\times \cases{[ (2+x^{2}+3\Delta_{s}^{2}/\Delta_{d}^{2})K(x)-
(2+2x^{2}+3\Delta_{s}^{2}/\Delta_{d}^{2})E(x)], & $x\le 1$
, \cr
x[ (1+2x^{2}+3\Delta_{s}/\Delta_{d}^{2})K(1/x)-
(2+2x^{2}+3\Delta_{s}^{2}/\Delta_{d}^{2})E(1/x)], 
& $x > 1$,} \nonumber
\end{eqnarray}
\begin{eqnarray}
\chi_{B_{2g}}^{\prime\prime \ sc.}&=&
\chi_{B_{2g}}^{\prime\prime}({\bf  q}=0,\omega)=\Theta(x^{2})
{4N_{F}\Delta_{d}\gamma_{B_{2g}}^{2}\over{3\pi \omega}} \\
&\times &
\cases{[ (1-x^{2})K(x)-(1-2x^{2}-3\Delta_{s}^{2}/\Delta_{d}^{2})E(x)], & 
$x\le 1$, \cr
x[ (2-2x^{2}-3\Delta_{s}^{2}/\Delta_{d}^{2}(1-1/x^{2}))K(1/x)-& \cr
(1-2/x^{2}-3\Delta_{s}^{2}/\Delta_{d}^{2})E(1/x)], 
& $x > 1$,} \nonumber
\end{eqnarray}
\begin{equation}
\chi_{A_{1g}}^{sc.}(i\omega)=\chi_{A_{1g},A_{1g}}(i\omega)-
\frac{\chi_{A_{1g},1}^2(i\omega)}{\chi_{1,1}(i\omega)},
\end{equation}
with the spectral functions
\begin{eqnarray}
& &\chi_{A_{1g},A_{1g}}^{\prime\prime}({\bf  q}=0,\omega)=\Theta(x^{2})
{4N_{F}\Delta_{d}\gamma_{A_{1g}}^{2}\over{15\pi \omega}} \\
&\times &\cases{
[ (7-8x^{2}+16x^{4}-5\Delta_{s}^{2}/\Delta_{d}^{2}(1-4x^{2}))K(x)- & \cr
(7-12x^{2}+32x^{4}-20\Delta_{s}^{2}/\Delta_{d}^{2}(1-2x^{2}))E(x)], & 
$x\le 1$, \cr
x[ (23-40x^{2}+32x^{4}+{5\Delta_{s}^{2}\over{\Delta_{d}^{2}x^{2}}}
(3-8x^{2}+8x^{4}))K(1/x)- & \cr
(11-28x^{2}+32x^{4}-{20\Delta^{2}_{s}\over{\Delta_{d}^{2}x^{2}}}(1-2x^{2}))
E(1/x)], & $x > 1$,} \nonumber
\end{eqnarray}
\begin{eqnarray}
\chi_{A_{1g},1}^{\prime\prime}({\bf  q}=0,\omega)&=&\Theta(x^{2})
{2\sqrt{2}N_{F}\Delta_{d}\gamma_{A_{1g}}\over{3\pi \omega}}  \\
&\times &\cases{
[ (1+2x^{2}+3\Delta_{s}^{2}/\Delta_{d}^{2})K(x)- & \cr
(1+4x^{2}+6\Delta_{s}^{2}/\Delta_{d}^{2})E(x)], & $x\le 1$, \cr
-x[ (1-4x^{2}+{3\Delta_{s}^{2}\over{\Delta_{d}^{2}x^{2}}}(1-2x^{2}))K(1/x)+
& \cr
(1+4x^{2}+6\Delta_{s}^{2}/\Delta_{d}^{2})E(1/x)], & $x > 1$,} \nonumber
\end{eqnarray}
and
\begin{equation}
\chi_{1,1}^{\prime\prime}({\bf  q}=0,\omega)=\Theta(x^{2})
{2N_{F}\Delta_{d}\over{\pi \omega}}
\cases{[ (1+\Delta_{s}^{2}/\Delta_{d}^{2})K(x)-E(x)],& $x\le 1$ , \cr
x[ (1+{\Delta_{s}^{2}\over{\Delta_{d}^{2}x^{2}}})K(1/x)-E(1/x)], & $x > 1$.}
\end{equation}
The results are plotted in Fig. (3) using an
value of $\Delta_{s}/\Delta_{d}=0.25$. The finite gap
$\Delta_{s}$ is responsible for the threshold appearing at
$2\Delta_{s}$, which is the minimum energy required to break a
Cooper pair. The spectra are polarization dependent, with the $B_{2g}$
and $A_{1g}$ spectra displaying a discontinuous jump at the
threshold while the $B_{1g}$ channel shows a continuous rise
from zero to a peak at 
$\omega=2\Delta_{max}=2\sqrt{\Delta_{s}^{2}+\Delta_{d}^{2}}$. 
Since both the $A_{1g}$ and $B_{2g}$ channels exhibit
broad maxima for the case of a pure $d_{x^{2}-y^{2}}$ gap, the presence
of the $2\Delta_{s}$ threshold will imply a shifting of the peak
of the spectra away from the pure case to values lower in
frequency if $2\Delta_{s} < \omega_{peak}$ (or at least create a
shoulder at $2\Delta_{s}$), while also removing
any difference in the peak position between the $A_{1g}$ and
$B_{2g}$ channels. Taken by itself this cannot be reconciled with the
experimental data unless of course $\Delta_{s}$ is very small.

\subsection{Anisotropic $s$--wave pairing}

Starting from band structure arguments, recently an anisotropic
$s$--wave energy gap of the form
\begin{equation}
\Delta({\bf \hat k})=\Delta_{0}+\Delta_{1}\cos^4(2\varphi)
\end{equation}
has been proposed also to explain the
cuprate materials \cite{sudip}. This energy gap is anisotropic
with predominantly $B_{1g}$--like character but does not change
sign around the Fermi surface. 
This is one specific example of an anisotropic $s-$wave energy gap. While
other representations for anisotropic $s-$wave gaps of course do exist
(in particular, a possibility recently suggested by photoemission
is $\Delta({\bf k})=\cos(k_xa)+\cos(k_ya) = const. + A \cos(4\varphi) 
\propto \cos^{2}(2\varphi)$\cite{photo}), we remark that the response
calculated here is not qualitatively different from other cases
and thus address only this one case. 

Again we evaluate the Raman response for such a superconductor
numerically and plot our results for
$\Delta_{0}/\Delta_{1}=0.25$ in Fig. (4). Immediately we
see similar behavior as in the previous case with one notable
exception. While the spectra each show a $2\Delta_{0}$
threshold, the $B_{2g}$ channel displays a
B.C.S.--like singularity at the threshold and the $A_{1g}$ a large increase
near the threshold that removes any
trace of a peak--like structure in the spectra at higher
frequencies. Again this would predict the same peak position (or a shoulder)
for the $A_{1g}$ and $B_{2g}$ channels. Therefore, for $\Delta_{0}$ not
too small, it is not possible 
using just the symmetry of the gap alone to arrive at a situation where the
peaks in the Raman spectra in the $A_{1g}$ and $B_{2g}$ channels
lie at separate high energies, again which is not in agreement 
with experiments.

In principle the $s$--wave component of the gap can be made vanishingly small. 
However, the greater anisotropy of the energy gap compared to the
$d_{x^{2}-y^{2}}$ case leads to a further anisotropy of the
position of the peaks in each channel, with peaks in the spectra
at $\omega=2\Delta_{max}, 0.6\Delta_{max},$ and
$0.2\Delta_{max}$ for the $B_{1g}, B_{2g},$ and $A_{1g}$
channels, respectively, where $\Delta_{max}=\Delta_{0}+\Delta_{1}$. 
On top of this, the low frequency
power--law behavior is linear in each channel, in contrast to the
$d_{x^{2}-y^{2}}$ results.  This is a general feature that the
spectra become more and more polarization dependent the greater the
anisotropy of the energy gap (compare to Fig. (1) for the $d_{x^{2}-y^{2}}$
case).

\section{Comparison with data on the cuprate systems and conclusions}

In this Section we present a comparison of the theory for a 
$d_{x^{2}-y^{2}}$ paired superconductor to recent measurements on the
electronic Raman continuum in three cuprate superconductors. 
In what follows, we plot $S(\omega)$, Eq. (1), 
which is given by $\chi$ of Eqs. (49)-(54)
mutliplied by the Bose factor.  In
drawing the fits to the spectra, the following procedure is employed.
First the fit to the $B_{1g}$ spectrum is made which determines the 
maximum value of the energy gap $\Delta_{0}$ via the position of the
peak in the spectrum. Next, the derived response is convoluted with
a Gaussian which mimics the effect of finite $z$--dispersion of the
Fermi surface, experimental resolution, inelastic scattering, etc.
Once this is done, the parameters remain fixed and only the prefactor of
the vertex is left to be adjusted to match the overall intensity, which
has no effect on the lineshape. Since $\gamma$ is in principle
derivable from band structure but presently unknown even for
such simple metals as Aluminum, this remains a free parameter. 

We first fit the data taken on single crystals of as grown 
Bi$_{2}$Sr$_{2}$CaCu$_{2}$O$_{8}$ (T$_{c}=90 K$) obtained in Reference
\cite{hackl1} for all symmetries at $T=20 K$, where a subtraction
procedure has been employed to ascertain the $A_{1g}$ signal 
(see Section 2.2). The comparison of the theory with experiment is
shown in Fig. (5). The parameters used to obtain the best fit to the
spectrum are $\Delta_{0}=287$ cm$^{-1}$ and a
smearing width of $\Gamma/\Delta_{0}=0.15$. The theory gives good agreement
with the data at low frequency shifts while at higher frequency shifts 
the theory fails to produce the broad continuum which is relatively
constant up to the scale of an $eV$.  This is most likely due to the
neglect of impurities and/or electron--electron scattering \cite{attila},
which is beyond the scope of the paper.
We see immediately that the peak
positions in the $B_{2g}$ and $A_{1g}$ channels given by the theory
automatially agree with the data. Also the asymptotic behavior of
the continuum at low frequencies given by Eq. (33) is shown in the data 
when one neglects the phonons at
roughly 100 and 330 wavenumbers. Again, these power--laws are intrinsic
to a $d_{x^{2}-y^{2}}$ pair state. Lastly, the ratio of the intensity
of the spectra in different channels is consistent with Eq. (35) and 
Fig. (2), which predicts that the $B_{1g}$ channels shows the smallest
intensity at low frequencies while the $A_{1g}$ channels shows the largest.
All of the experimental features are thus consistent with the theory at 
least at low frequency shifts $\omega < 1000$ cm$^{-1}$.  

We now turn to the data taken on single crystals of YBa$_{2}$Cu$_{3}$O$_{7}$
(T$_{c}=88 K$) obtained in Ref. \cite{hackl} for all symmetries at $T=20 K$,
where the same subtraction procedure used in BSCCO was employed to
ascertain the $A_{1g}$ signal. The comparison of the theory to the data
is shown in Fig. (6), with the parameters $\Delta_{0}=210$ cm$^{-1}$ and 
$\Gamma/\Delta_{0}=0.2$ Again, the theory gives a good description of the
data for $\omega < 1000$ cm$^{-1}$.
We again see the peak positions at relatively
the same place as in BSCCO, and power--laws linear in frequency at low
shifts for the $B_{2g}$ and $A_{1g}$ channels. We note that while the
cubic rise of the spectra predicted by the theory fits rather well with
the $B_{1g}$ data in BSCCO, we remark that the Fano effect of 
the $B_{1g}$ phonon
which appears to be stronger in YBCO than in BSCCO can obscure the 
rise of the spectra at low frequencies at give the
appearance of a linear dependence on frequency \cite{hung}. Lastly, the
ratios of the response in the static limit again are consistent with the
theory.

Lastly, we investigate the single layer Thallium compound
Tl$_{2}$Ba$_{2}$CuO$_{6}$ (T$_{c}= 80 K$) obtained in Ref.
\cite{hackl3}. The sample is most likely
the most affected by disorder and therefore our theory will not be
expected to give the best fit to the data. Our fits is shown in Fig. 7
for the $B_{1g}$ and the mixed $A_{1g}+B_{2g}$ channels at $T=20 K$,
where we have ignored the contribution from the $B_{2g}$ channel since it
is believed to be minor compared to $A_{1g}$. The value of the parameters
used are $\Delta_{0}=232 {\rm cm}^{-1}$ and smearing width 
$\Gamma/\Delta_{0}=0.25$.  All phonons have been subtracted.
Once again the theory gives a good description of the relative peak 
positions.  Considering also that this compound has only one Cu-O layer,
the agreement of the theory with experiment also validates the assumption
that the Raman scattering results predominantly from  
intraband fluctuations of the single Cu-O layer band, and that interband
scattering can be neglected.
The theory also accounts for the
linear rise of the spectrum for the
$A_{1g}$ channel for low frequencies. However, the theory cannot account
for the linear rise of the spectrum in the $B_{1g}$ channel. This most
likely is due to the neglect of impurity scattering.
The residual scattering near zero frequency shifts is also
borne out by the theory, although the amount is underestimated for the
$B_{1g}$ channel. Again, this most likely has to do with the neglect of
impurity scattering.

We remark that the theory is incomplete in that the theory fails
to describe the flat continuum at large frequency shifts. Moreover, the
theory cannot be extended to the normal state since the response functions
vanish at $T_{c}$ in the limit $q \rightarrow 0$ due to phase space
restrictions. Here the additional
physics of electron--electron scattering and/or impurity scattering must be 
incorporated to have a consistent theory to simultaneously describe the
normal and superconducting state data. This remains a topic of further
research \cite{imp}. Nevertheless, the low frequency behavior of the
spectra, and in particular, the relative peak positions of each polarization
channel are quantitatively described by the theory.

Thus we have seen that the Raman measurements on the cuprate
systems provide a large body of symmetry dependent information all of
which agrees with the predictions of $d_{x^{2}-y^{2}}$ pairing.  Of
course at present the information from Raman alone cannot completely
rule out the
possibility of the presence of a very small gap which exists over the
entire Fermi surface nor can it determine whether the gap changes sign
around the Fermi surface. However, the theoretical comparison shows
that the gap must be predominantly of $B_{1g}$ character, and the low
temperature and low frequency data seem to indicate that the $s$--wave
component of the gap must be very small if it exists at all. More precise
measurements could of course clarify this point further. Also, more work 
is needed from band structure to pin down magnitude of the Raman tensor 
elements to predict the overall intensities. 

\acknowledgments

The authors would like to thank Drs. R. Hackl, G. Krug, R. Nemetschek and
B. Stadlober for providing us with their data and discussions.
Similarly, we thank D. Reznik, R. T. Scalettar,
A. Virosztek, A. Zawadowski, and G. T. Zimanyi for enlightening
discussions.
This work was supported in part by N.S.F. Grant number 92-06023
and by the American Hungarian Joint Grant Number NSF 265/92b. 
One of the authors (T.P.D.)
would like to acknowledge the hospitality of the Walther
Meissner Institute, the Research Institute for Solid State
Physics of the Hungarian Academy of Science, and the Institute
of Physics of the Technical University of Budapest
where parts of this work were completed. 

\appendix{Weak coupling results}

In this appendix we want to show that a gap anisotropy causes both the gap at
zero temperature $\Delta_0(0)$ and the specific heat discontinuity at the
transition $\Delta C/C_{\rm N}$ (related to the slope of the gap function
near $T_{\rm c}$) to deviate from their respective BCS values of
$(\Delta_0(0)/T_{\rm c})_{\rm BCS}=\pi\exp(-\gamma)=1.7638...$ and 
$(\Delta C/C_{\rm N})_{\rm BCS}=12/7\zeta(3)=1.4261...$, with
$\gamma=.57721...$ and $\zeta(3)=1.20205...$ denoting Euler's constant and
Riemann's zeta function respectively. 
A straightforward solution of Eq. (18) leads to 
\begin{eqnarray}
\frac{\Delta_0(0)}{T_{\rm c}}&\equiv&\delta_{\rm sc}=
\left(\frac{\Delta_0(0)}{T_{\rm c}}\right)_{\rm BCS}\exp\left(
-\frac{<|\Delta_p|^2
\ln(|\Delta_p|/\Delta_0)>_{\rm FS}}{<|\Delta_p|^2>_{\rm FS}}\right),
\nonumber \\
\frac{\Delta C}{C_{\rm N}}&=&
\left(\frac{\Delta C}{C_{\rm N}}\right)_{\rm BCS}
\frac{<|\Delta_p|^2>_{\rm FS}^2}{<|\Delta_p|^4>_{\rm FS}}.
\end{eqnarray}  
These results may be used to generate an interpolation formula for the
temperature dependence of the gap maximum $\Delta_0(T)$:
\begin{equation}
\Delta_0(T)=\delta_{\rm sc} T_{\rm c}
\tanh\left[\frac{\pi}{\delta_{\rm sc}}\sqrt{\frac{3}{2}
\frac{\Delta C}{C_{\rm N}}
\frac{\Delta_0^2}{<|\Delta_p|^2>_{\rm FS}}\left(\frac{T_{\rm c}}{T}-1\right)}
\right].
\end{equation}
For the special case of a gap with $d_{x^2-y^2}$ symmetry, one obtains for
the parameters $\delta_{\rm sc}$ and $\Delta C/C_{\rm N}$ to following
numbers:
\begin{eqnarray}
\delta_{\rm sc}&=&\cases{\frac{\pi}{2}\exp\left(-\gamma+\frac{16}{15}\right)=
2.5626, & \ \ \ \ {\rm spherical\ FS,} \cr
2\pi\exp\left(-\gamma-\frac{1}{2}\right)=               
2.1397, & \ \ \ \ {\rm cylindrical\ FS,}}\nonumber \\
\Delta C/C_{\rm N}&=&\cases{\frac{12}{7\zeta(3)}\frac{7}{15}=
0.6655, & \ \ \ \ {\rm spherical\ FS,} \cr
\frac{12}{7\zeta(3)}\frac{2}{3}=
0.9507, & \ \ \ \ {\rm cylindrical\ FS.}}\nonumber
\end{eqnarray}
We should emphasize, that these numbers should not be taken too seriously
since they emerge from a weak coupling treatment. One should rather adopt the
convention to treat them as parameters which can be adjusted to experiment 
and so account for strong coupling effects in the trivial sense in which
they appear as renormalizations of the quantities $\Delta_0$ and $
\Delta C/C_{\rm N}$.

\appendix{impurity scattering in the normal state}

In this Appendix we would like to demonstrate that the normal state
electronic Raman response has a nonvanishing ${\bf q}\rightarrow 0$ limit
the presence of quasiparticle collisions.  
In the collisionless limit of a superconductor, the quantity 
$\chi_{\gamma\gamma}-\Delta_{\gamma\gamma}$ has a 
finite value in the normal state limit $\Delta_0\rightarrow 0$ 
only at finite wavenumbers ${\bf q}$. The normal state Raman response can
persist in the limit ${\bf q}\rightarrow 0$ however if one assumes 
(elastic or
inelastic) scattering processes to take place. These scattering processes 
may be generally introduced through a collision integral 
$\delta I_k$ in the normal state kinetic equation\cite{PN},
\begin{eqnarray}
\omega\delta n_k&-&{\bf q}\cdot{\bf v}_k \ h_k=i\delta I_k\{h_k\},
\nonumber \\
\delta I_k\{h_k\}&=&-\frac{1}{\tau_k}h_k+2\sum_{\bf p}\ C_{k p}\ h_p, \\
h_k&=&\delta n_k-\frac{\partial n_k^0}{\partial \xi_k}\delta\epsilon_k. 
\nonumber
\end{eqnarray}
The structure of the collision integral depends on the scattering
mechanism that one is interested in and is, in many cases, subject to an
approximate treatment of the collision operator
$C_{k p}$. However, such approximations have to be consistent
with the requirement of charge conservation \cite{NDM,EH,DE}.
An approximation with this property is the  
separate kernel approximation \cite{EC},
\begin{equation}
C_{k p}=\left(-\frac{\partial n_k^0}{\partial \xi_k}\right)\
\lambda_0\ \frac{1}{\tau_k}\frac{1}{\tau_p}
\left\{2\sum_{p}\left(-\frac{\partial n_p^0}{\partial \xi_p}\right)
\frac{1}{\tau_p}\right\}^{-1}.
\end{equation}
The scattering processes are here represented by a momentum--dependent
quasiparticle lifetime $\tau_k$ and a scattering parameter $\lambda_0$.
The continuity equation is obtained from Eq. (B1) by summing over momenta
${\bf k}$, 
$$
\omega \delta n-{\bf q}\cdot{\bf j}=
-2 i \sum_{\bf p}\frac{h_p}{\tau_p}(1-\lambda_0).
$$
Particle number conservation is therefore connected with the
choice of the scattering parameter $\lambda_0=1$. One may show that the 
normal state Raman response in the ${\bf q}\rightarrow 0$ limit assumes form
equivalent to the result (26) for the pair breaking Raman effect: 
\begin{eqnarray}
\lim_{\Delta_0\rightarrow 0}\lim_{{\bf q}\rightarrow 0}\chi_{\gamma\gamma}&=&
-\left\{\Lambda_{\gamma\gamma}
-\lambda_0\ \frac{\Lambda_{\gamma 1}^2}{\Lambda_{1 1}}\right\}, \nonumber \\
\Lambda_{a b}&=&2\sum_{\bf p}\left(
-\frac{\partial n_k^0}{\partial \xi_k}\right)
\frac{a_p b_p}{1-i\omega\tau_p} \ \ .
\end{eqnarray}
Like in the superconducting case, the projected structure of Eq. (B3) 
has its origin in the particle number conservation law. This result
can be applied to impurity or phonon scattering as well as to inelastic
two-particle scattering and requires the specification of the
momentum-dependent quasiparticle lifetime for each of these cases.

\appendix{Diagrammatic Gauge Invariant Raman Response: role of vertex 
corrections and collective modes}

In this Appendix we derive expressions for gauge--invariant
generalized correlation functions using a diagrammatic approach
with the main emphasis placed
on electronic Raman scattering. After solving the coupled
integral equations for the renormalized vertex, we evaluate the 
position, broadening and
residue of the massive and massless
collective modes as a function of coupling
strength for a $d_{x^{2}-y^{2}}$ energy gap on a spherical Fermi
surface. We then investigate the collective modes on a cylindrical
(2-D) Fermi surface by turning off any $z$--dispersion in the
band structure.

        We begin by writing down the expression for the general two-particle
response function in Nambu space as
\begin{equation}
\chi({\bf  q},i\omega)=-T\sum_{i\omega_n}\sum_{{\bf  k}} Tr[\hat\Gamma({\bf 
k, q},i\omega) \hat G({\bf  k}+\frac{{\bf q}}{2},i\omega_n)
\hat\gamma({\bf  k, -q}) \hat G({\bf  k}-\frac{{\bf q}}{2},i\omega_n-i\omega)],
\end{equation}
\begin{eqnarray}
\hat\Gamma({\bf  k, q},i\omega)-\hat\gamma({\bf  k, q})=
-T\sum_{i\omega_n}\sum_{{\bf  p,
p^{\prime}}} \tau_{3}\hat G({\bf  p+\frac{p^{\prime}}{2}},i\omega_{n}) \hat
\Gamma({\bf  p,p^{\prime}},i\omega) \\ \nonumber
\times\hat G({\bf  p-\frac{p^{\prime}}{2}},i\omega-i\omega_{n}) 
V({\bf  p-k}+{{\bf  q-p^{\prime}}\over{2}},
{\bf  k-p}+{{\bf  q-p^{\prime}}\over{2}})\tau_{3},
\end{eqnarray}
where $Tr$ denotes taking the trace, $\tau_{i}$ are Pauli matrices in
Nambu space, and the vertex $\hat\gamma$
determines the correlation function of interest.  
The dressed vertex $\hat\Gamma$
contains the interactions $V$ responsible for maintaining gauge invariance.

Our consideration is
focused on the Raman vertex which describes the anisotropic mass
fluctuations, $\hat\gamma({\bf  k,q})=
\tau_{3}\gamma({\bf k})$ with the limit of small ${\bf  q}$ taken.
Other choices of the vertices are
\begin{eqnarray}
\gamma({\bf  k})&=&\tau_{3},~~~~~~ \hfil {\rm charge~~~density,} \nonumber \\
\gamma({\bf  k})&=&\tau_{0}, ~~~~~~ \hfil {\rm spin~~~density,}  \\
\gamma({\bf  k})&=&{\bf  k}\tau_{0}, ~~~~~~ \hfil {\rm current~~~density}. 
\nonumber
\end{eqnarray}
Thus from Eqs. (C3) 
we see that while the spin and charge density vertices probe only the
$L=0$ channel and current density
probes $L=1$, in principle all even $L$ channels
contribute to the Raman vertex. In fact the charge density is the first term
in the expansion of the Raman vertex.

        If one replaces the dressed vertex $\Gamma$ by the undressed one,
Eq. (2), then of course Eq. (C1) is manifestly not gauge invariant, and in
general the neglect of collective modes arising from a gauge
invariant treatment could in principle
affect the overall spectrum.  
Usually the question of gauge invariance is
rather an academic one since the modes that appear in BCS systems have
little impact on the response functions of a superconductor.
It is well known that due to the spontaneously broken $U(1)$ gauge symmetry in
$s$--wave superconductors, two collective modes appear -- an optical one
with a frequency of $2\Delta$ which is damped -- and a sound--like mode, the
Anderson--Bogolubov mode, which is soft and lies in the gap for neutral
superconductors but is raised to the plasmon energy by the long range
Coulomb forces via the Higgs mechanism.  However, in unconventional 
superconductors, there
can exist in principle additional Goldstone modes corresponding to the 
additional broken continuous
symmetries such as $SO_{3}^{S}$ spin rotational symmetry in
spin--triplet systems plus $SO_{3}^{L}$ orbital rotational symmetry in
spin--singlet systems if the gap does not possess the full symmetry of
the Fermi surface. In addition, massive collective modes can
arise if the energy gap is degenerate or has an admixture of
different representations from the point group. The
massive modes can in principle lie below the gap edge
and thus be relevant for the low frequency dynamics of correlation functions.
For example, the spectrum of collective modes in
$^{3}$He is well known in both phases and lead to
observable effects \cite{wolfle}.
However, the collective modes of possible
$d$--wave states that might be candidates for strongly correlated systems
are not as well understood \cite{hira}. There are indications
that a $d$--wave state of $d_{x^{2}-y^{2}}$ is particularly favorable in
systems with strong correlations \cite{theory}, thus underscoring the
necessity of an understanding of the response functions and collective
modes for such a superconductor.

We continue our calculation for the gauge invariant Raman response 
by first expanding the renormalized matrix vertex
$\hat\Gamma$ along the Fermi surface in terms of crystal harmonics,
\begin{equation}
\hat\Gamma({\bf  \hat k},i\omega)=\sum_{L,m}\hat\Gamma_{L}^{m}(i\omega)
\phi_{L}^{m}({\bf  \hat k}),
\end{equation}
and do the same for the pairing interaction,
\begin{equation}
V({\bf  \hat k, \hat p})=\sum_{L,L^{\prime};m,m^{\prime}}
V_{L,L^{\prime}}^{m,m^{\prime}}\phi_{L}^{m}({\bf  \hat k})
\phi_{L^{\prime}}^{m^{\prime}}({\bf  \hat p}).
\end{equation}
The integral equation for the renormalized vertex at ${\bf  q}=0$
can then be written as
\begin{eqnarray}
\hat\Gamma^{m}_{L}(i\omega)-\hat\gamma^{m}_{L}&=& 
-{TN_{F}\over{2}}\sum_{i\omega_n}\int d\epsilon
d\epsilon^{\prime}
\sum_{L^{\prime},L^{\prime\prime};m^{\prime},m^{\prime\prime}}
V_{L,L^{\prime}}^{m,m^{\prime}} \\ 
&\times & \langle\phi_{L^{\prime\prime}}^{m^{\prime\prime}}({\bf  \hat k})
\phi_{L^{\prime}}^{m^{\prime \*}}({\bf  \hat k})
\tau_{3}\hat G({\bf  \hat k},i\omega_n-i\omega)
\hat\Gamma_{L^{\prime\prime}}^{m^{\prime\prime}}(i\omega)
\hat G({\bf  \hat k},i\omega_n)\tau_{3}\rangle, \nonumber
\end{eqnarray}
where $\langle \cdots \rangle$ denotes an average over the Fermi surface
and $N_{F}$ denotes the density of states at the Fermi level. 
Eq. (C6) is completely general for any type of vertex, interaction and
gap symmetry. In particular, for the case of an isotropic energy gap,
Eq. (C6) recovers the previous results for the density, current \cite{big}, and
Raman reponses \cite{klein}. We confine ourselves to the case of singlet
energy gaps and use the BCS approximation,
\begin{equation}
\hat G({\bf k},i\omega_n)={i\omega_n+\xi\tau_{3}+
\Delta({\bf  \hat k})\tau_{1}\over{(i\omega_n)^{2}-E^2({\bf k})}},
\end{equation}
where $E^2({\bf k})=\xi^2+\Delta^2({\bf  \hat k})$. More complicated
Green's functions could treated with the same scheme. However, the analysis
gets considerably more complicated and cannot be carried as far analytically.

In general the pairing interaction $V$ can have off--diagonal as well as
diagonal terms in the $L$ basis, and in general all channels
will be coupled.  If the interaction has the symmetry of the
Fermi surface then the integral equations only couple different 
channels $L$ and $L^{\prime}$ which transform
according to the same irreducible representation. However, the
subsequent matrix can be diagonalized with respect to the
indices of the same representation resulting in a new set of basis
functions which are linear combinations of the old basis
functions of the same representation in different $L$ channels \cite{itai}.
Thus we can then write the interaction as a diagonal matrix in
the new basis functions which still has a general structure for each 
representation within each new channel. Thus we can write 
$V_{L,L^{\prime}}^{m,m^{\prime}}=
V_{L}^{m}\delta_{L,L^{\prime}}\delta_{m,m^{\prime}}$. This
allows us to reduce the infinite
series of coupled integral equations to a limited subset that
can be handled analytically. The
elements of the expansion will be dominated by a single $V_{L}^{m}$
component corresponding to the $L$ pairing channel symmetry $m$, and the
other components represent admixtures of channels (the 
smaller eigenvalues in the gap equation) with different pairing
symmetry. If the gap representation is one dimensional (all
representations of the $D^{4h}$ group except $E_{g}$, which is 
two dimensional), then all the other $V_{L}^{m}$'s are set to zero and
there will only be collective modes connected with the broken
$U(1)$ gauge invariance. Otherwise other collective modes can be
present as well.

We now define new vertices
$\hat\Gamma_{L}^{m}(i\omega)=\hat\gamma_{L}^{m}+
V_{L}^{m}\hat\delta_{L}^{m}(i\omega)$ and
expand $\hat\delta_{L}^{m}(i\omega)$ in spin quaternions,
\begin{equation}
\hat\delta_{L}^{m}(i\omega)=\sum_{i=0}^{3}\delta^{(i)m}_{L}(i\omega)\tau_{i}.
\end{equation}
The index $m$ stands for
the representations of the $D^{4h}$ point group, and are defined through
the basis functions as follows
\begin{equation}
\phi_{L=2}^{m}({\bf \hat k})=\cases{
{1\over{\sqrt{2}}}(k_{x}^{2}-k_{y}^{2}), &$B_{1g} (\Gamma_{3}^{+}) \ (m=1),$\cr
{1\over{\sqrt{6}}}(2k_{z}^{2}-k_{x}^{2}-k_{y}^{2}), &$A_{1g} (\Gamma_{1}^{+}) \ (m=2),$\cr
k_{x}k_{y}, &$B_{2g} (\Gamma_{4}^{+}) \ (m=3),$\cr
k_{x}k_{z}, &$E_{g} (\Gamma_{5}^{+})\ (m=4),$\cr
k_{y}k_{z}, &$E_{g} (\Gamma_{5}^{+})\ (m=5),$}
\end{equation}
where the $\Gamma_{i}^{+}$ corresponds to the notation of Sigrist and Rice
\cite{sigrist}. 
It can be shown that the coefficients $\delta^{(0,1)m}_{L}$ of
Pauli matrices $\tau_{0,1}$ satisfy homogeneous equations and thus
vanish while the remaining coefficients satisfy the following
coupled integral equations,
\begin{equation}
\delta_{L}^{(2)m}(i\omega)=
\sum_{L^{\prime};m^{\prime}}\bigl\{V_{L^{\prime}}^{m}
\delta_{L^{\prime}}^{(2)m}(i\omega) C^{-m,m^{\prime}}_{L,L^{\prime}}(i\omega)+
i(\gamma_{L^{\prime}}^{m^{\prime}}+V_{L^{\prime}}^{m^{\prime}}
\delta_{L^{\prime}}^{(3)m^{\prime}}(i\omega))
A_{L,L^{\prime}}^{m,m^{\prime}}(i\omega)\bigr\},
\end{equation}
\begin{equation}
\delta_{L}^{(3)m}(i\omega)=\sum_{L^{\prime};m^{\prime}}
\bigl\{iV_{L^{\prime}}^{m^{\prime}}
\delta_{L^{\prime}}^{(2)m^{\prime}}(i\omega) 
A^{m,m^{\prime}}_{L,L^{\prime}}(i\omega)-
(\gamma_{L^{\prime}}^{m}+V_{L^{\prime}}^{m}
\delta_{L^{\prime}}^{(3)m}(i\omega))
C_{L,L^{\prime}}^{+m,m^{\prime}}(i\omega)\bigr\},
\end{equation}
where the functions $A, C^{\pm}$ are given by
\begin{eqnarray}
A_{L,L^{\prime}}^{m,m^{\prime}}(i\omega)&=&
\langle \phi_{L}^{m\*}({\bf  \hat k})
\phi_{L^{\prime}}^{m^{\prime}}({\bf  \hat k})
A({\bf  \hat k},i\omega)\rangle,\nonumber \\
C_{L,L^{\prime}}^{\pm m,m^{\prime}}
(i\omega)&=&\langle \phi_{L}^{m\*}({\bf  \hat k})
\phi_{L^{\prime}}^{m^{\prime}} ({\bf  \hat k})
C^{\pm}({\bf  \hat k},i\omega)\rangle.
\end{eqnarray}
The spectral functions are defined as
\begin{eqnarray}
A({\bf  \hat k},i\omega)&=&-i\Delta({\bf \hat k})
\omega N_{F}\int d\xi {1\over{4E^{2}}}
{1-2f(E)\over{i\omega-2E}}- (i\omega \rightarrow -i\omega),\nonumber \\
C^{+}({\bf  \hat k},i\omega)&=&-N_{F}\int d\xi
{\Delta({\bf  \hat k})^{2}\over{2E^{2}}}
{1-2f(E)\over{i\omega-2E}}+ (i\omega \rightarrow -i\omega),\nonumber \\
C^{-}({\bf  \hat k},i\omega)&=&-{N_{F}\over{2}}\int d\xi
{1-2f(E)\over{i\omega-2E}} + (i\omega \rightarrow -i\omega).
\end{eqnarray}
Here $f$ is a Fermi function
and $(i\omega \rightarrow -i\omega)$ denotes additional terms which differ only
in the sign of $i\omega$. Analytically continuing to the real axis
by letting $i\omega \rightarrow \omega +i0$, the $\xi$ integration
can be performed analytically and we obtain
\begin{eqnarray}
A({\bf  \hat k},\omega)&=& \Delta({\bf \hat k})\ \ F({\bf  \hat k},\omega), \nonumber \\
C^{+}({\bf  \hat k},\omega)&=& {2\Delta^2({\bf  \hat k})\over{\omega}}
F({\bf  \hat k},\omega),\ \\
C^{-}({\bf  \hat k},\omega)&=& {1\over{V({\bf  \hat k})}} +
{\omega\over{2}}F({\bf  \hat k},\omega), \nonumber
\end{eqnarray}
with
\begin{equation}
F({\bf  \hat k},\omega)= \cases{
{N_{F}\over{\sqrt{\Delta({\bf  \hat k})^{2}-(\omega/2)^{2}}}}\arctan
\left[{\omega\over{2\sqrt{\Delta({\bf  \hat k})^{2}-(\omega/2)^{2}}}}\right],
& {\rm for}\ \ $\Delta({\bf  \hat k})^{2}>({\omega\over{2}})^{2}$, \nonumber \cr
{N_{F}\over{2\sqrt{(\omega/2)^{2}-\Delta({\bf  \hat k})^{2}}}}\left(i\pi+
\log\left[{\omega/2-\sqrt{(\omega/2)^{2}-\Delta({\bf  \hat
k})^{2}}\over{\omega/2+\sqrt{(\omega/2)^{2}-\Delta({\bf  \hat k})^{2}}}}\right]
\right)
,& {\rm for}\ \ $\Delta({\bf  \hat k})^{2}\le ({\omega\over{2}})^{2}$,}
\end{equation}
and
\begin{equation}
{1\over{V({\bf  \hat k})}}=N_F
\int_{0}^{\infty}{d\xi\over{2 E({\bf k})}}
\end{equation}
given by the BCS gap equation. The function $F$ is closely related to the
Tsuneto function, Eq. (11).

The integral equations are still general for
a charge density--like vertex and the symmetry of the interaction and gap
remain undefined. We now restrict our attention to the case of $d$--wave
interactions such that only $V_{L=2}^{m} \ne 0$ and other terms
corresponding to interactions in higher angular momentum channels
are discarded. Dropping the $L=2$ subscript by denoting $V_{L=2}^{m}$
by $V_{m}$ and $\delta_{L=2}^{(2,3)m}$ by $\delta_{m}^{(2,3)}$,
the integral equations simplify to
\begin{equation}
\delta^{(2)}_{m}(i\omega)=\sum_{m^{\prime}}\bigl\{V_{m^{\prime}}
[\delta_{m^{\prime}}^{(2)}(i\omega) C_{L=2,L=2}^{- m,m^{\prime}}(i\omega)+
i\delta_{m^{\prime}}^{(3)}(i\omega)A_{L=2,L=2}^{m,m^{\prime}}(i\omega)]+
i\sum_{L^{\prime}}\gamma_{L^{\prime}}^{m^{\prime}}A_{L=2,L^{\prime}}^{m,
m^{\prime}}(i\omega)\bigr\},
\end{equation}
\begin{equation}
\delta_{m}^{(3)}(i\omega)=\sum_{m^{\prime}}\bigl\{
V_{m^{\prime}}[-\delta_{m^{\prime}}^{(3)}(i\omega)
C_{L=2,L=2}^{+m,m^{\prime}}(i\omega)
+i\delta_{m^{\prime}}^{(2)}(i\omega) A^{m,m^{\prime}}_{L=2,L=2}(i\omega)]-
\sum_{L^{\prime}}\gamma_{L^{\prime}}^{m}C^{+ m,m^{\prime}}_{L=2,L^{\prime}}
(i\omega)\bigr\},
\end{equation}
These equations are still general to any $d$--wave pair state.

We now specifically work with a $B_{1g}$ gap,
$\Delta({\bf  \hat k})=\Delta_{0}(\hat k_{x}^{2}-\hat k_{y}^{2})$ (the
$\Gamma_{3}^{+}$ representation \cite{sigrist}), noting
that similar conclusion can be drawn for other choices of energy gaps
within the $L=2$ subgroup of the $D^{4h}$ point group. The coupled
integral equations represent 10 equations for the 10 unknowns
$\delta^{(i)}_{m}, i \in \{2,3\}$, and the solution can be obtained
by diagonalizing a $10 \times 10$ matrix. 
In aiding to solve the coupled integral equations, it is
useful to examine the selection rules of the spectral functions
$A, C^{\pm}$ for various subgroup indices within the $L=0,2$ channels. In
particular we note that
\begin{eqnarray}
C^{\pm m,m^{\prime}}_{L,L}\sim \delta_{m,m^{\prime}}, \ \ \ \ \
C^{\pm 4,4}_{L=2,L=2}=C^{\pm 5,5}_{L=2,L=2},\ \ \ \ \
C^{\pm 0,m}_{L=0,L=2} \sim \delta_{m,2},
\nonumber \\
A_{L=2,L=2}^{m,m^{\prime}} \sim \delta_{m,m^{\prime}}, {\rm for}\ m,m^{\prime}
\notin \{2,3\}, \ \ \ \ \
A_{L=2,L=2}^{1,m} \sim \delta_{m,2},\ \ \ \ \
A_{L=2,L=2}^{2,m} \sim \delta_{m,1}, \nonumber \\
A_{L=0,L=0}^{0,0}=0, \ \ \ \ \  A_{L=2,L=0}^{m,0} \sim \delta_{m,1}, \ \ \ \ \
A_{L=2,L=2}^{4,4}=-A_{L=2,L=2}^{5,5}.
\end{eqnarray}
These theorems hold for the case of a $B_{1g}$ gap only.

Handling the full gauge
invariant response amounts to solving a matrix equation
to identify the collective modes in each channel.  Solving these
equations allows one to identify the position of the collective modes
by locating the zeroes of the real part of the denominator.  We first
simplify our notation by defining
\begin{eqnarray}
f^{A}_{m}(i\omega)&=&\sum_{L,m^{\prime}}\gamma_{L}^{m^{\prime}}
A_{L=2,L}^{m,m^{\prime}}(i\omega),
\nonumber \\
f^{C}_{m}(i\omega)&=&-\sum_{L,m^{\prime}}\gamma_{L}^{m^{\prime}}
C_{L=2,L}^{+m,m^{\prime}}(i\omega).
\end{eqnarray}
We find that the $m=1 (B_{1g})$ and $m=2 (A_{1g})$ channels are coupled
due to the fact that a $d_{x^{2}-y^{2}}$ energy gap squared has a
component which has a finite overlap with the $A_{1g}$ channel which is
isotropic within the $x-y$ plane. Solving the integral equations we obtain
for the $m=1,\ (B_{1g})$ and $m=2,\ (A_{1g})$ channels,
\begin{eqnarray}
\delta_{1}^{(2)}(i\omega)&=&i {f_{1}^{A}(i\omega)+V_{2}A_{2,2}^{1,
2}(i\omega)\delta_{2}^{(3)}(i\omega)\over{1-V_{1}C^{-1,1}_{2,2}(i\omega)}},
\nonumber \\
\delta_{1}^{(3)}(i\omega)&=&{f_{1}^{C}(i\omega)-V_{2}{f_{2}^{A}(i\omega)A_{2,
2}^{1,2}(i\omega)\over{1-V_{2}C_{2,2}^{-2,2}(i\omega)}}\over{1+V_{1}C_{2,2}^{+1,
1}(i\omega)+{V_{1}V_{2}\mid A_{2,2}^{1,2}(i\omega)\mid^{2}\over{1-V_{2}C_{2,
2}^{-2,2}(i\omega)}}}},\nonumber \\
\delta_{2}^{(2)}(i\omega)&=& i{f_{2}^{A}(i\omega)+V_{1}A_{2,2}^{1,
2}(i\omega)\delta_{1}^{(3)}(i\omega)\over{1-V_{2}C_{2,2}^{-2,2}(i\omega)}},
\nonumber \\
\delta_{2}^{(3)}(i\omega)&=&{f_{2}^{C}(i\omega)-V_{1}{f_{1}^{A}(i\omega)A_{2,
2}^{1,2}(i\omega)\over{1-V_{1}C_{2,2}^{-1,1}(i\omega)}}\over{1+V_{2}C_{2,2}^{+2,
2}(i\omega)+{V_{1}V_{2}\mid A_{2,2}^{1,2}(i\omega)\mid^{2}\over{1-V_{1}C_{2,
2}^{-1,1}(i\omega)}}}}.
\end{eqnarray}
For the $m=3 (B_{2g})$ and $m=4,5$ (two $E_{g}$) channels, we find no coupling
and obtain
\begin{eqnarray}
\delta_{3}^{(2)}(i\omega)&=& i {f_{3}^{A}(i\omega)\over{1-V_{3}C_{2,2}^{-3,
3}(i\omega)}}, \nonumber \\
\delta_{3}^{(3)}(i\omega)&=& {f_{3}^{C}(i\omega)\over{1+V_{3}C_{2,2}^{+3,
3}(i\omega)}},
\end{eqnarray}
\begin{eqnarray}
\delta_{4,5}^{(2)}(i\omega)=i{f_{4,5}^{A}(i\omega)+V_{4,5}A_{2,2}^{4,4;5,
5}(i\omega)\delta_{4,5}^{(3)}(i\omega)\over{1-V_{4,5}C_{2,2}^{-4,4;5,
5}(i\omega)}},
\nonumber \\
\delta_{4,5}^{(3)}(i\omega)={f_{4,5}^{C}(i\omega)-V_{4,5}{f_{4,5}^{A}(i\omega)
A_{2,2}^{4,4;5,5}(i\omega)\over{1-V_{4,5}C_{2,2}^{-4,4;5,
5}(i\omega)}}\over{1+V_{4,5}C_{2,2}^{+4,4;5,5}(i\omega)+{\mid V_{4,5}A_{2,2}^{4,
4;5,5}(i\omega)\mid^{2}\over{1-V_{4,5}C_{2,2}^{-4,4;5,5}(i\omega)}}}},
\end{eqnarray}
where $V_{4,5}, \delta_{4,5}, C^{\pm 4,4;5,5},$ and $A^{4,4;5,5}$
stand for the interactions, renormalized vertices,
and spectral functions in the $m=4,5$ channels, respectively.  These
equations represent the full channel dependent renormalization of the
Raman vertex due to $d-$wave pairing interactions.

We now are in a position to identify the collective modes in the 10 channels
(5 real and 5 imaginary) by locating the zeroes of the denominator in each
channel. While the functions $f^{A,C}$ remain general and are determined by
the magnitude of the Raman vertex they do not enter into the denominators of
the expressions for the renormalized vertices and thus are irrelevant for
the position and broadening of the collective modes. The modes located
by the zeroes of the denominator are of two types, namely, massive and
massless. The massless (Goldstone) modes are a consequence of the spontaneously
broken $U(1)$ continuous symmetry by the interactions, while the
massive (optical) modes are generated as well. Using Eqs. (C21-C23),
in each channel we 
locate the zeroes of the real part of the denominator in each channel to find
the position of the collective mode and evaluate the imaginary part of the
denominator at the position of the collective mode to determine its
broadening. 

Our results are summarized in Table 1.  We see that the Goldstone
mode ($\omega_{c}=0$) appears in the $A_{1g}$ channel and the other modes
are all massive. The $B_{1g}$ modes lie beneath the maximum energy gap
while the others appear above $2\Delta_{0}$. These excitonic modes are
damped considerably due to the existence of quasiparticles from the presence
of gap nodes which provide decay channels to damp the massive
modes. This is in contrast to excitonic modes in $s$--wave superconductors
\cite{klein,me}. However, we do observe in Table 1 that the mode position 
rapidly decreases to lower frequencies with depreciating residue as the
coupling stregnth $V_{m}/V_{B_{1g}}$ is reduced, which is similar to
the BCS case. Below a
critical coupling strength $V_{m}/V_{B_{1g}} \sim 0.8$ the
collective modes disappear altogether and thus have little
impact on the low frequency behavior of the Raman correlation
function for small couplings.
 
We now reconstruct the Raman response and determine which of the collective
modes couple to the Raman vertex. Putting in Eqs. (C21)-(C23) into (C1), we
can express the full gauge invariant Raman response function as
\begin{eqnarray}
& &\chi_{L}^{m}({\bf  q}=0,i\omega)= \\
& &2\gamma_{L}^{m}\sum_{m^{\prime}}
(\sum_{L^{\prime}}\gamma_{L^{\prime}}^{m^{\prime}}
C_{L,L^{\prime}}^{+m,m^{\prime}}(i\omega)
+V_{m^{\prime}}\delta_{m^{\prime}}^{(3)}(i\omega)
C_{L,2}^{+m,m^{\prime}}(i\omega)-
iV_{m^{\prime}}\delta_{m^{\prime}}^{(2)}(i\omega)
A_{L,2}^{m,m^{\prime}}(i\omega)). \nonumber
\end{eqnarray}
Taking only the $L=0, 2$ terms of the Raman vertex into account (see
Section III), we carry out the summation over $L^{\prime}$ and $m^{\prime}$
to obtain the Raman spectrum in each channel, and then 
take into account the long--range Coulomb screening via \cite{klein}
\begin{equation}
\chi_{sc}({\bf  q} \rightarrow 0,i\omega)=
\chi_{\gamma,\gamma}(i\omega)-
\frac{\chi_{\gamma,\gamma_{L=0}}^2(i\omega)}
{\chi_{\gamma_{L=0},\gamma_{L=0}}(i\omega)}\ ,
\end{equation}
where $\chi_{A,B}$ denotes Eq. (C1) with the vertices $A({\bf  \hat
k})$ and $B({\bf  \hat k})$, respectively.  

We find that for the
density channel ($L=0$), $\chi({\bf  q}=0,\omega)=0$, which is a restatement of
particle number conservation and bears out the gauge invariant nature
of the theory. This simply restates that intercell charge fluctuations 
couple to the long--range Coulomb forces to be completely screened for
$q \rightarrow 0$. After lengthy
but trivial algebra we obtain the compact results for the intracell
fluctuation contributions. For the $B_{1g}$ channel we obtain,
\begin{equation}
\chi_{B_{1g}}(i\omega)=2(\gamma_{2}^{1})^{2}
{C_{B_{1g}}(i\omega)\over{1+V_{1}C_{B_{1g}}(i\omega)}},
\end{equation}
and for the $B_{2g}$ channel,
\begin{equation}
\chi_{B_{2g}}(i\omega)=2(\gamma_{2}^{3})^{2}
{C_{B_{2g}}(i\omega)\over{1+V_{3}C_{B_{2g}}(i\omega)}},
\end{equation}
and for the two $E_{g}$ channels,
\begin{equation}
\chi_{E_{g}}(i\omega)=2(\gamma_{2}^{4,5})^{2}
{C_{E_{g}}(i\omega)\over{1+V_{4,5}C_{E_{g}}(i\omega)}},
\end{equation}
with the functions
\begin{equation}
C_{B_{1g}}(i\omega)=C^{+1,1}_{2,2}(i\omega)+V_{2}{\mid A_{2,2}^{1,2}(i\omega)
 \mid^{2}\over{1-V_{2}C^{-2,2}_{2,2}(i\omega)}},
\end{equation}
\begin{equation}
C_{B_{2g}}(i\omega)=C^{+3,3}_{2,2}(i\omega),
\end{equation}
\begin{equation}
C_{E_{g}}(i\omega)=C^{+4(5),4(5)}_{2,2}(i\omega)+V_{4(5)}{\mid A_{2,2}^{4(5),
4(5)}(i\omega)
 \mid^{2}\over{1-V_{4(5)}C^{-4(5),4(5)}_{2,2}(i\omega)}}.
\end{equation}
These channels are unaffected by screening since the intracell
fluctuations lead to no net charge transfer for these symmetries. 
The expression for the fully symmetric $A_{1g}$ channel which contains 
both inter--and intracell fluctuations expression is
much more complicated due to Coulomb screening,
\begin{equation}
\chi_{A_{1g}}(i\omega)=2\gamma_{2}^{2}[C_{2,2}^{+2,2}(i\omega)- 
C_{2,0}^{+2,0}(i\omega)^{2}/C_{0,0}^{+0,0}(i\omega)]{1-
{\gamma_{0}\over{\gamma_{2}}}
V_{2}C_{A_{1g}}^{0}(i\omega)\over{1+V_{2}C_{A_{1g}}(i\omega)}}.
\end{equation}
Here the functions $C_{A_{1g}}, C_{A_{1g}}^{0}$ are defined as
\begin{equation}
C_{A_{1g}}(i\omega)=C_{2,2}^{+2,2}(i\omega)+V_{1}
{\mid A_{2,2}^{1,2}(i\omega)\mid^{2}\over{1-V_{1}C^{-1,1}_{2,2}(i\omega)}},
\end{equation}
and
\begin{equation}
C_{A_{1g}}^{0}(i\omega)=C_{2,0}^{+2,0}(i\omega)+V_{1}
{A_{2,2}^{1,2}(i\omega)A_{2,0}^{1,0}(i\omega)\over{1-V_{1}C^{-1,1}_{2,
2}(i\omega)}}.
\end{equation}

Examing the structure of the denominator of Eqs. (C26)-(C31), we see that the
massless gauge mode (Goldstone mode) in the $A_{1g}$ channel drops out
of the Raman response, and only provides screening via Eq. (C25).
The massive modes
do appear in the $B_{1g}$ and $E_{g}$ channels while the massive mode
found in the $B_{2g}$ channel does not couple to a Raman probe.
Linearizing the denominator around the position of the collective mode we
calculate the residue $Z$ of the mode to be
$Z_{B_{1g}}=\omega_{c}/\log(\Delta_{0}/\omega_{c})$ and $Z_{E_{g}}=\omega_{c}$,
where $\omega_{c}$ is the position of the collective mode in each channel.
Therefore, putting all our results together for the collective modes, we
can argue that the collective modes are of little relevance to electronic
Raman scattering due to the fact that 1)the modes only exist at
beyond a large strong couplings threshold, 2) the residue of the modes 
decreases the lower the position of the collective mode, and 3)
that the broadening of the collective modes are substantial.

However, of course the final--state interactions themselves can
affect the spectrum \cite{klein,me}. Therefore we display our results in 
Figs. (8a-c) for
the entire spectrum in the $B_{1g}, B_{2g},$ and  $A_{1g}$ channels
for different ratios of the parameter $V_{m}/V_{B_{1g}}$, where $V_{B_{1g}}$ 
is the pairing interaction of $d_{x^{2}-y^{2}}$ symmetry. 
The response for the $E_{g}$ channels are similar to the $B_{2g}$ case.
We note that the
interactions have only a minor affect on the spectra in the $B_{1g}$ channel,
only changing the cusp behavior near $2\Delta_{0}$, (this demonstrates
how the finite
$z$--dispersion cuts off the logarithmic singularity in Fig. 1 for the
2--D Fermi surface), while leaving the peak
position and low frequency behavior unchanged. The interactions have more
an effect in the $B_{2g}$ and $A_{1g}$ channels due to the fact that the 
peak of the
spectra are very broad. Therefore the interactions shift the peak
position upwards in frequency along the top of the broad 
hump of the spectrum, from $1.3$ to $1.5 \Delta_{0}$ for the $B_{2g}$ channel,
and from $0.6$ to $0.8\Delta_{0}$ in the $A_{1g}$ channel.
Again we note that the low frequency behavior
remains unchanged. A similar effect is seen in the $E_{g}$ channels. In
particular, the massive mode in this channel is entirely damped leading to
no drastic changes in the spectrum.

We close this Appendix by addressing the collective modes for a
superconductor with a cylindrical (2--D) Fermi surface. In
a 2--D system the only interactions that appear at the $L=2$ level
are the $B_{1g}$ and $B_{2g}$ channels, while the $A_{1g}$ and
$A_{2g}$ channels appear at $L=4$. Similarly, there are no
$E_{g}$ channels for a system without dispersion in the
$z$--direction. Therefore, if we are only concerned with $d$--wave
interactions and not those of higher angular momentum channels,
then we can set the matrix elements
$V_{A_{1g}}=V_{E_{g}(1,2)}=0$ in Eqs. (C21-C23) and the equations
simplify greatly. Since then there
will be no channel mixing of the $A_{1g}$ channels into the
response, then therefore there can be no collective mode since
no term appears in the denominators which has the form
$1-V_{m}/V_{B_{1g}}$ which arises through the function $C^{-}$
(see Eqs. (C21-C23)). Therefore we can conclude that collective modes
can only appear when the interactions are included in higher
order $L$ channels for a 2--D system. Furthermore, the vertex corrections
produce minor changes in the spectra only for the $B_{1g}$ and $B_{2g}$
channels. The changes are similar to the changes shown for the response
functions evaluated on a spherical Fermi surface (apart from cutting off
the logarithmic divergence of the $B_{1g}$ response at the gap edge), 
and thus are not of major importance.
We can therefore neglect the
collective modes entirely and the effect of vertex corrections and
simply use the bare bubble for the
Raman response. This completes the purpose of the Appendix.


\figure{Electronic Raman response functions evaluated for $d_{x^{2}-y^{2}}$
pairing on a cylindrical
Fermi surface for various symmetries as indicated. All vertices have been
set equal to 1.}

\figure{Ratio of the low frequency Raman response in the superconductor
with $d_{x^{2}-y^{2}}$ pairing
to the normal metal for the temperatures indicated. Note the slow decrease
of the $A_{1g}$ and $B_{2g}$ channels with temperature.}

\figure{Response functions for various symmetries for $s+id$ pairing
with $\Delta_{s}/\Delta_{d}=0.25$. All vertices are set to 1.}

\figure{Response functions for various symmetries for 
anisotropic $s$--wave pairing
with $\Delta_{0}/\Delta_{1}=0.25$. All vertices are set to 1.}

\figure{Comparison of the theory to the experimental data taken on
BSCCO from Ref. \cite{hackl1} using $d_{x^{2}-y^{2}}$ pairing. 
The parameters used are defined in the text.}

\figure{Comparison of the theory to the experimental data taken on
YBCO using $d_{x^{2}-y^{2}}$ pairing from Ref. \cite{hackl}. 
The parameters used are defined in the text.}

\figure{Comparison of the theory to the experimental data taken on
single layer Thallium cuprate 
using $d_{x^{2}-y^{2}}$ pairing from Ref. \cite{hackl3}. 
The parameters used are defined in the text.}

\figure{Effect of vertex corrections on the Raman response evaluated 
for $d_{x^{2}-y^{2}}$ pairing on a spherical Fermi surface for
(a) $B_{1g}$, (b) $B_{2g}$, and (c) $A_{1g}$ channels (using
$\gamma_{0}/\gamma_{2}=2$, other vertices set equal to 1). The $E_{g}$
results look identical to the $B_{2g}$ spectra. Here we have used
$V_{B_{1g}}=0.2$, and the values of $V_{m}/V_{B_{1g}}$ are indicated in
the upper right hand corner of the Figure.}

\newpage
\epsffile{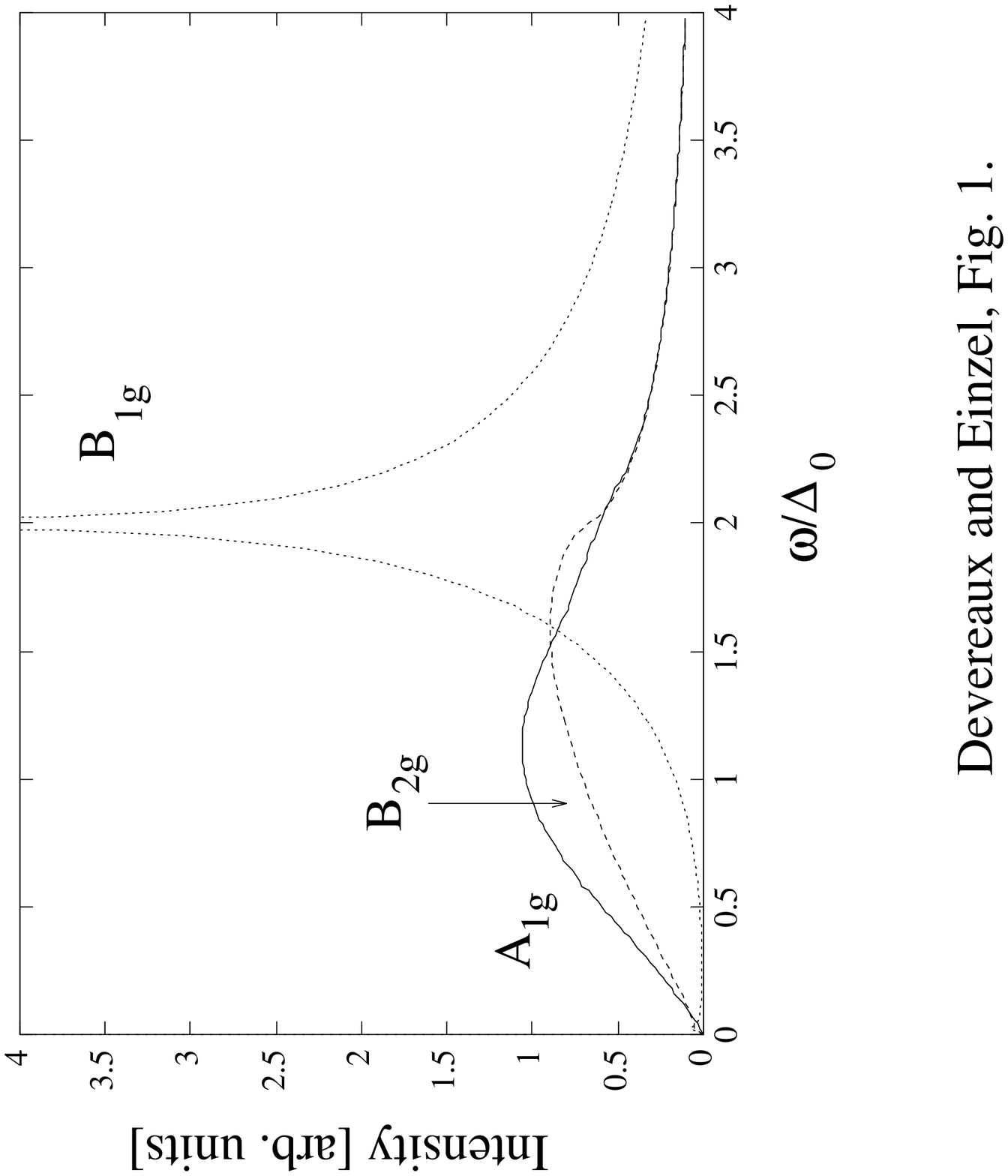}
\newpage
\epsffile{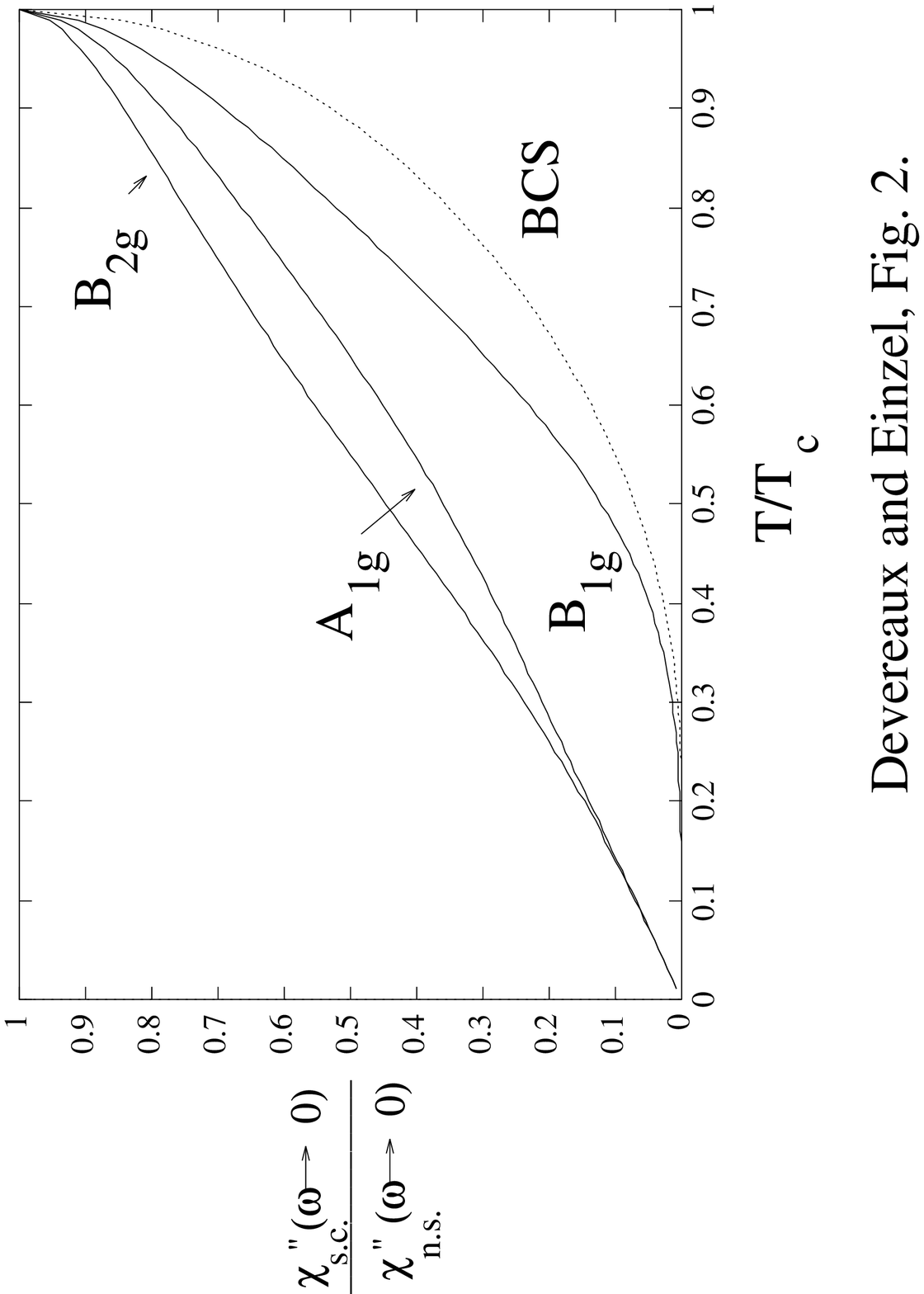}
\newpage
\epsffile{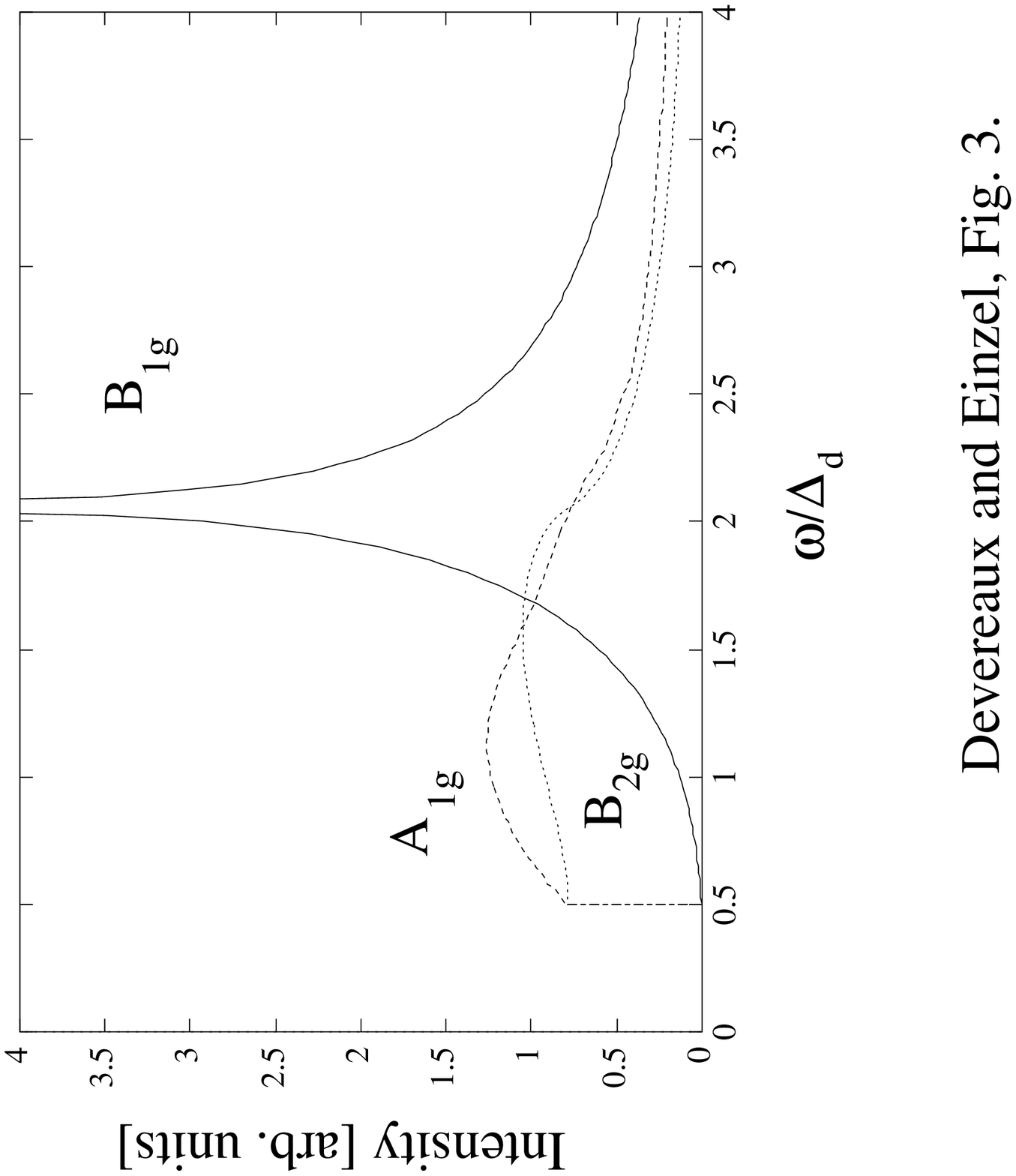}
\newpage
\epsffile{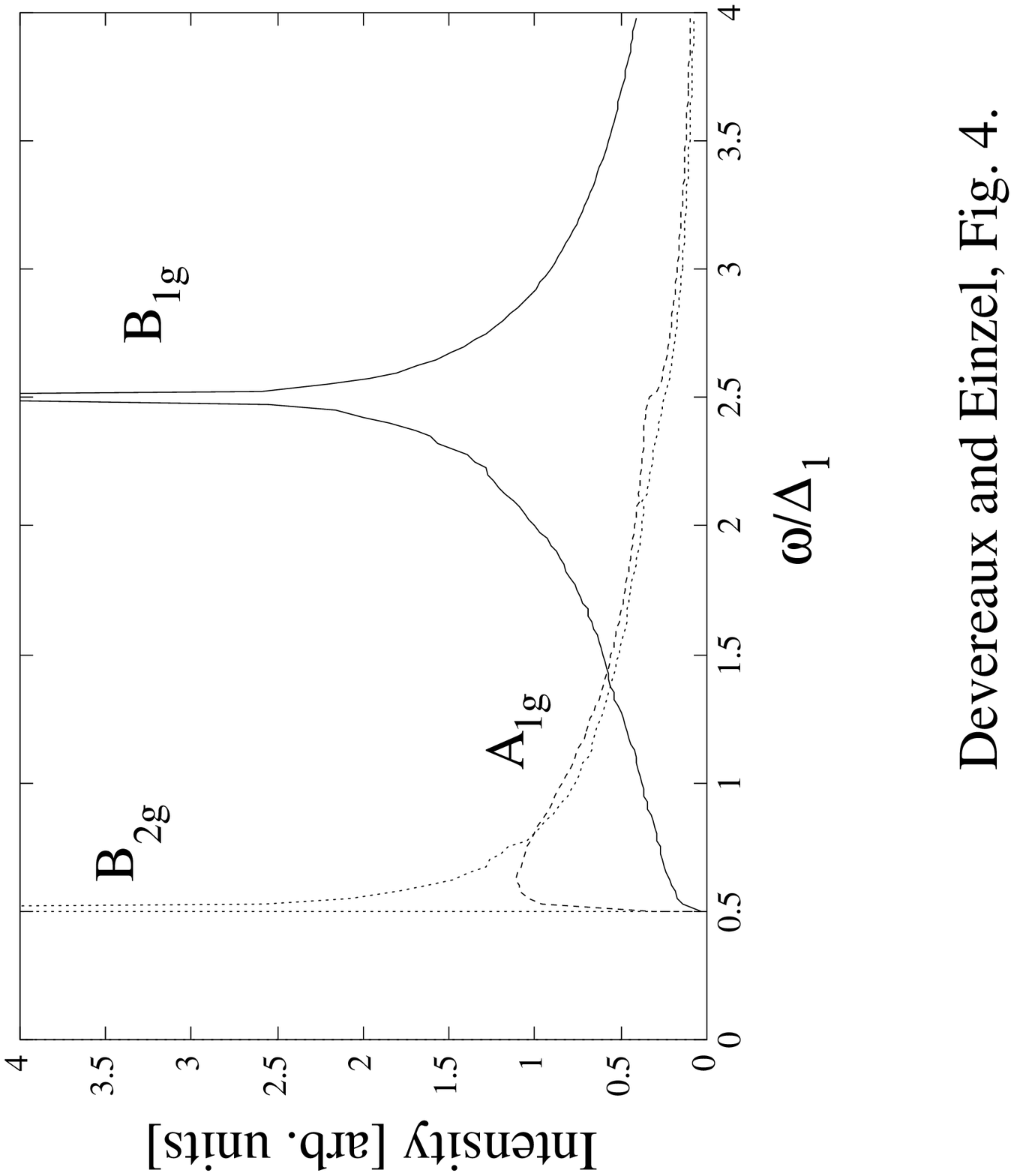}
\newpage
\epsffile{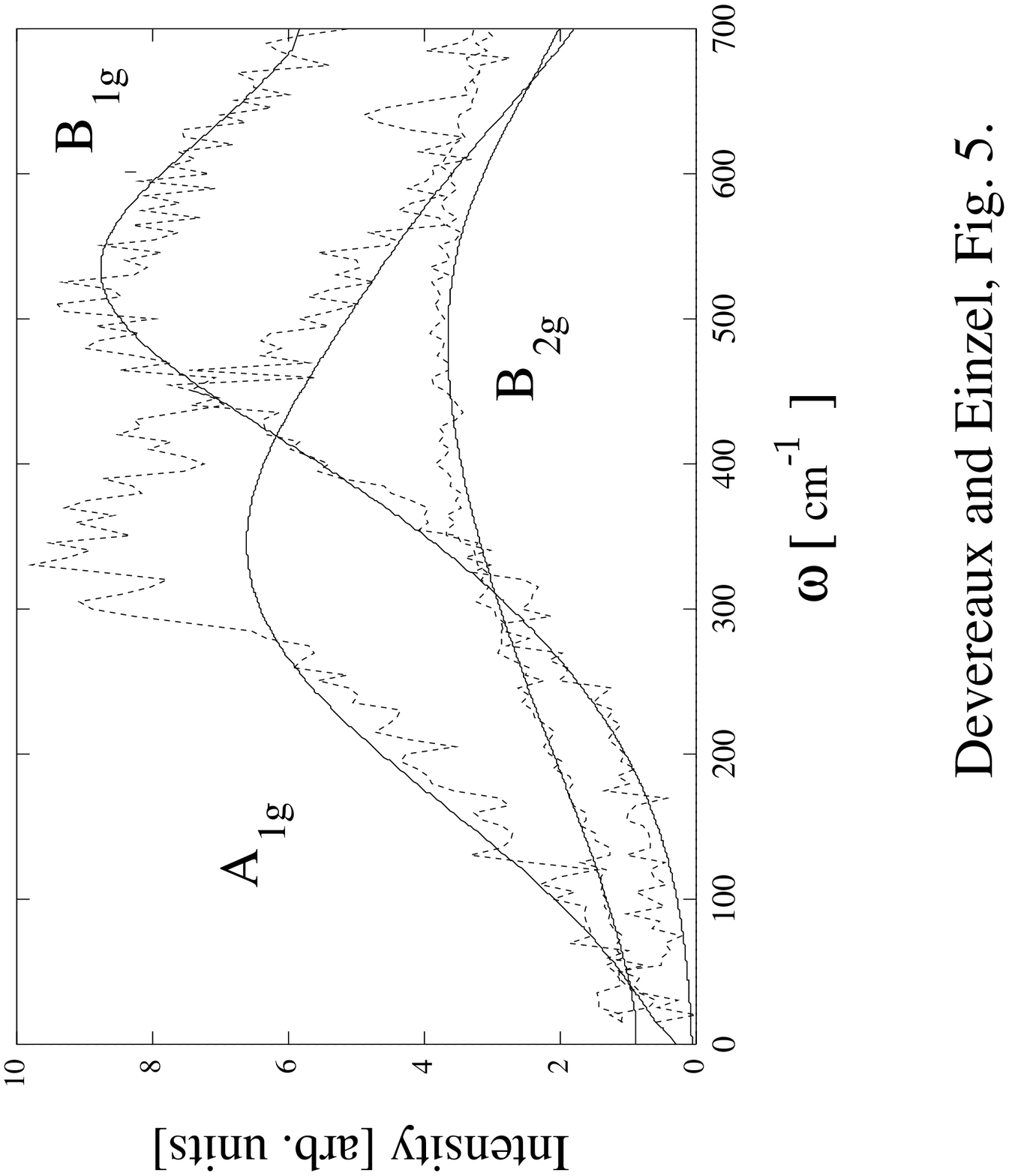}
\newpage
\epsffile{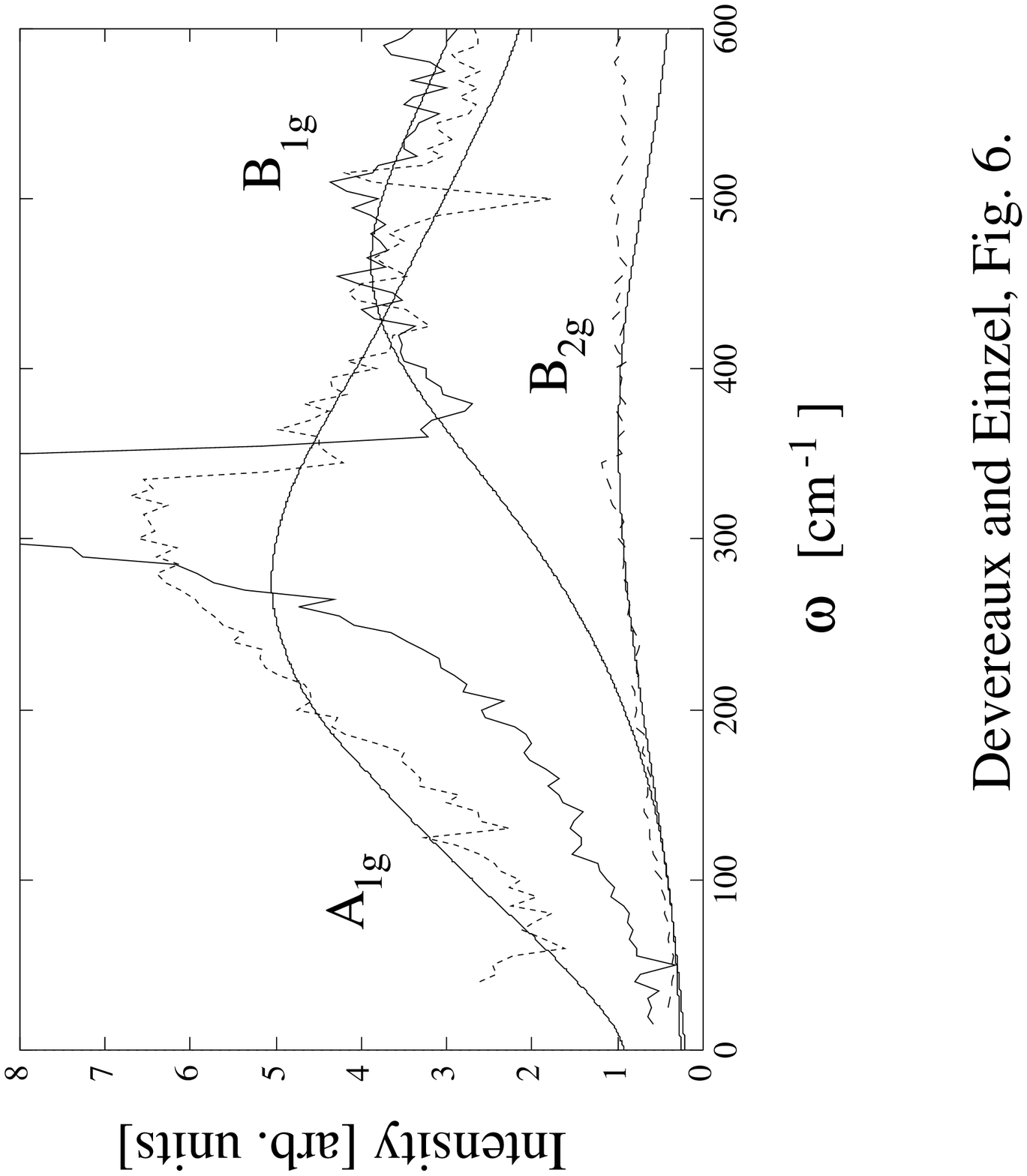}
\newpage
\epsffile{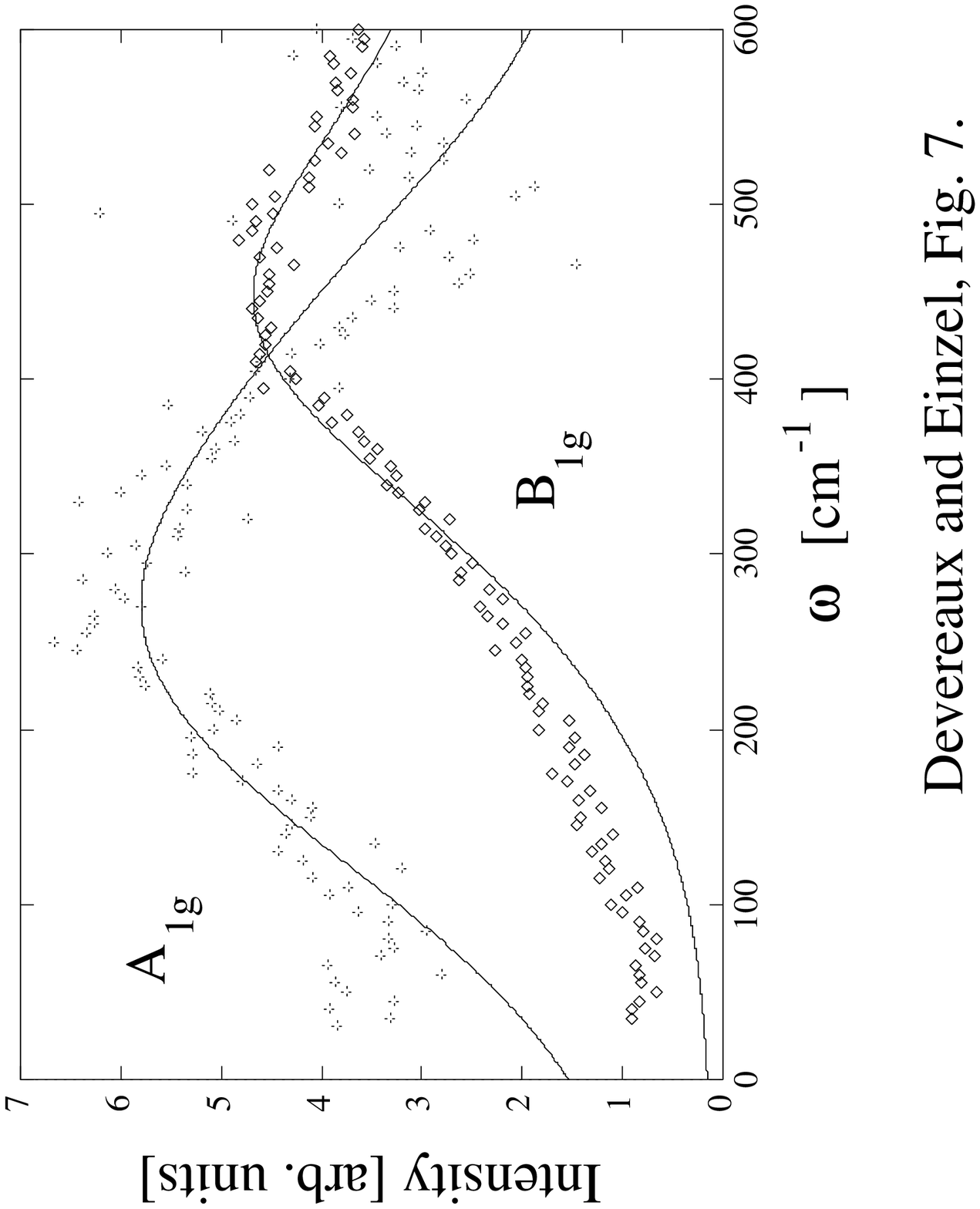}
\newpage
\epsffile{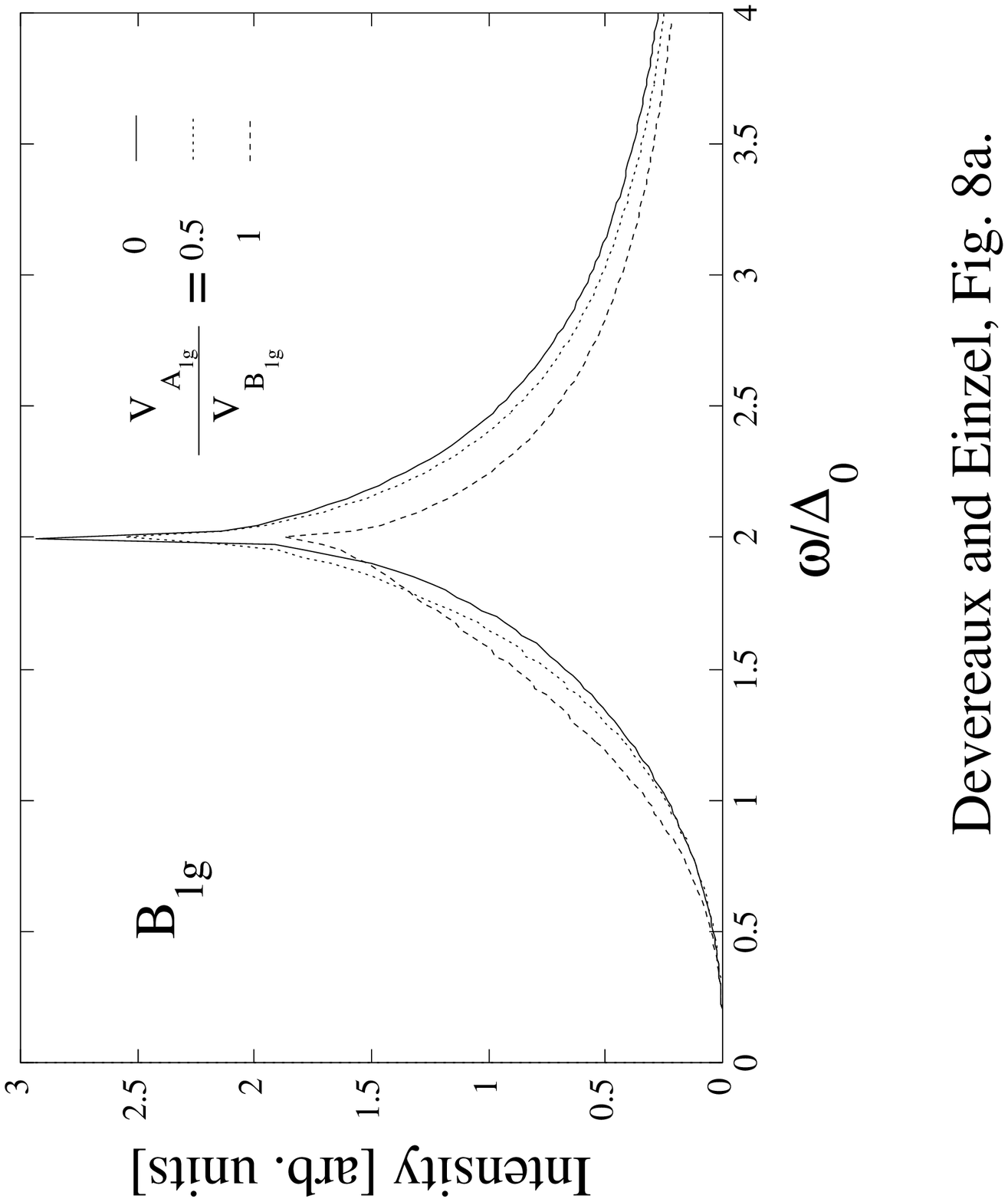}
\newpage
\epsffile{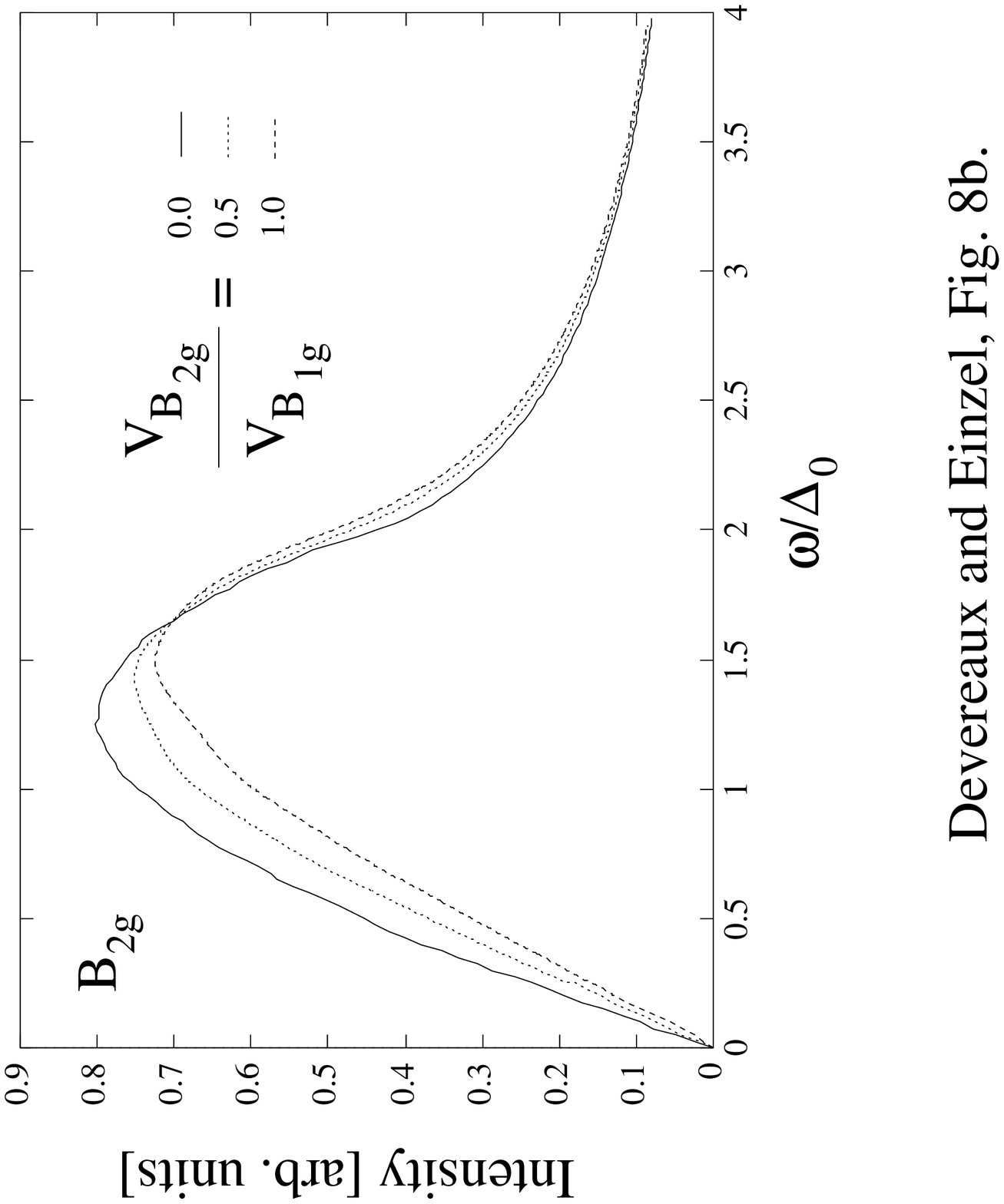}
\newpage
\epsffile{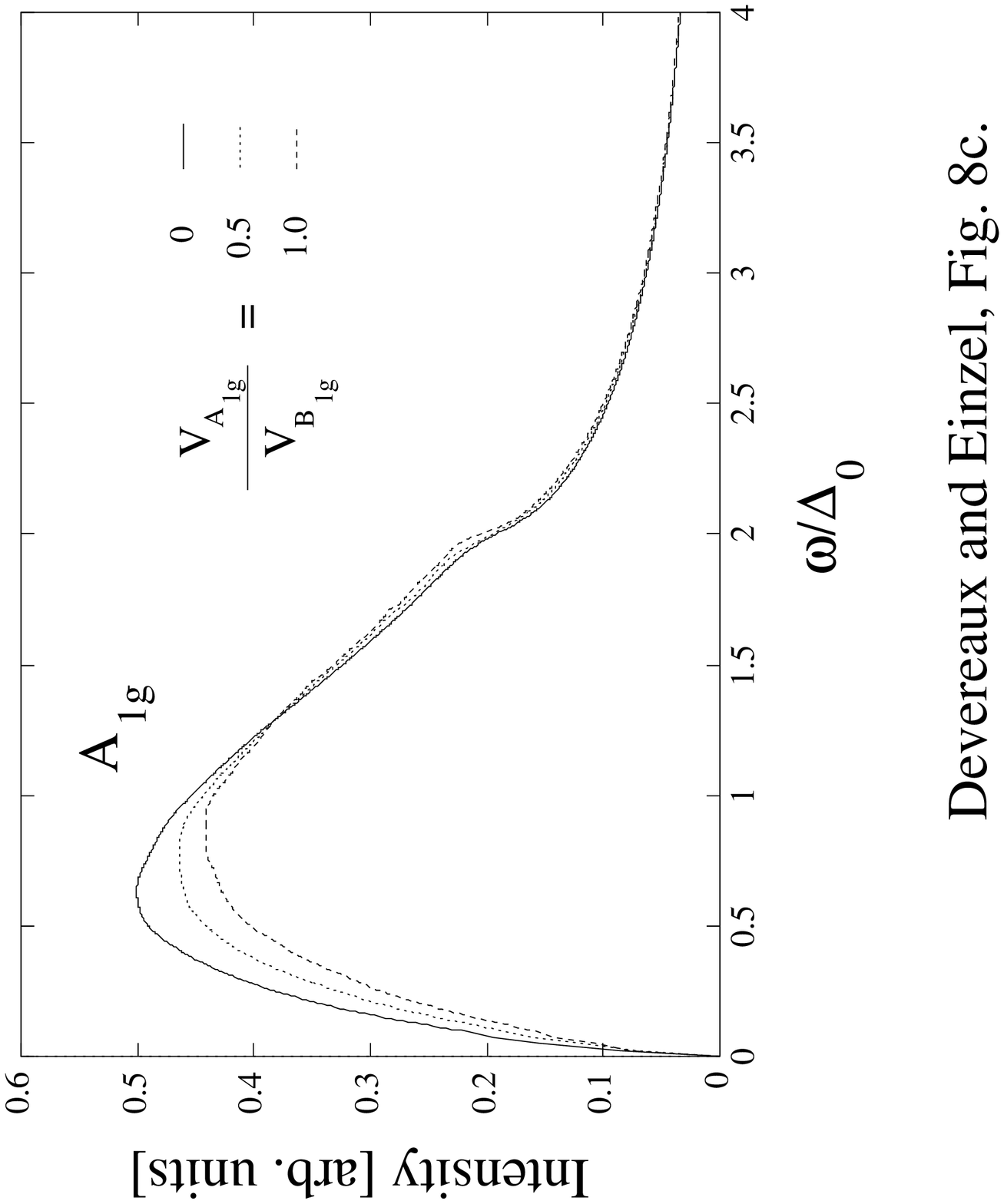}
\newpage
\centerline{TABLE 1}

\centerline{Position $\omega_{c}/2\Delta_{0}$ and broadening 
$\Gamma_{c}/2\Delta_{0}$ of the pole in $\delta^{(2,3)}_{m}$ 
for each channel.}
\vskip 1cm
\begin{tabular}{|l|l|l|l|l|l|l|l|l|}\hline\hline
--- & $\delta^{(2)}_{1}\ (B_{1g})$ & $\delta^{(3)}_{1}\ (B_{1g})$ &
$\delta^{(2)}_{2}\ (A_{1g})$ & $\delta^{(3)}_{2}\ (A_{2g})$ & $\delta^{(2)}_{3}\
 (B_{2g})$ & $\delta^{(3)}_{3}\ (B_{2g})$ & $\delta^{(2)}_{4,5}\ (E_{g})$ &
$\delta^{(3)}_{4,5}\ (E_{g})$ \cr \hline
${\omega_{c}\over{2\Delta_{0}}}$ & \ \ \ 0.87 & \ \ \ 0.83 & Goldstone & Goldstone & \ \
 \ 1.16& \ \ \ 1.16 & \ \ \ 1.15& \ \ \ 1.09 \cr \hline
${\Gamma_{c}\over{2\Delta_{0}}}$ & \ \ \ 0.17 & \ \ \ 0.20 & \ \ \ \ --- & \ \ \ \ --- &
 \ \ \ 0.23 & \ \ \ 0.23 & \ \ \ 0.23 & \ \ \ 0.22 \\ \hline
\end{tabular}


\begin{references}
\bibitem{better}
This indicates that the energy gap can only be identified with the
peak position in the $B_{1g}$ channel, and the temperature dependence
of the peak position will most closely follow its true form only in
this channel. However, it is to be noted that impurities will also
change the position of the
peak away from $2\Delta_{0}$ even in this channel as well (see Ref.
\cite{me}), making the identification of the gap from the peak maximum less
precise.
\end{references}
\end{document}